\documentclass[final, 5p,times]{elsarticle}

\usepackage{booktabs} % For formal tables
\usepackage{graphicx}
\usepackage{subfigure}
\usepackage{amsmath}
\usepackage[dvipsnames]{xcolor}
\usepackage{amssymb, array, listings}
 
 \usepackage[outdir=./img/]{epstopdf}
 \graphicspath{{img/}}

\usepackage{url}
\usepackage{hyperref}
\usepackage{color}

\lstdefinelanguage{XML}
{
	morestring=[b]",
	morestring=[s]{>}{<},
	morecomment=[s]{<?}{?>},
	stringstyle=\color{black},
	identifierstyle=\color{blue},
	keywordstyle=\color{cyan},
	morekeywords={xmlns,version,type}% list your attributes here
}

\journal{Parallel Computing}
\begin{document}
\begin{frontmatter}
\title{Checkpoint/Restart Approaches for a Thread-Based MPI Runtime}

\author[1]{Julien Adam}
\author[2]{Maxime Kermarquer}
\author[1]{Jean-Baptiste Besnard}
\author[3]{Leonardo Bautista-Gomez}
\author[2]{Marc P\'erache}
\author[2]{Patrick Carribault}
\author[2]{Julien Jaeger}
\author[4]{Allen D. Malony}
\author[4]{Sameer Shende}
	
\address[1]{ParaTools SAS, Bruy\`eres-le-Ch\^atel, France}
\address[2]{CEA, DAM, DIF, F-91297 Arpajon, France}
\address[3]{Barcelona Supercomputing Center, Barcelona, Spain}
\address[4]{ParaTools Inc., Eugene, United States }

\begin{abstract}

    Fault-tolerance has always been an important topic when it comes to running
    massively parallel programs at scale.  Statistically, hardware and software
    failures are expected to occur more often on systems gathering millions of
    computing units.  Moreover, the larger jobs are, the more computing hours
    would be wasted by a crash.  In this paper, we describe the work done in
    our MPI runtime to enable both transparent and application-level
    checkpointing mechanisms.  Unlike the MPI 4.0 User-Level Failure Mitigation
    (ULFM) interface, our work targets solely Checkpoint/Restart and ignores
    other features such as resiliency.  We show how existing checkpointing
    methods can be practically applied to a thread-based MPI implementation
    given sufficient runtime collaboration. The two main contributions are the
    preservation of high-speed network performance during transparent C/R and
    the over-subscription of checkpoint data replication thanks to a dedicated
    user-level scheduler support. These techniques are measured on MPI
    benchmarks such as IMB, Lulesh and Heatdis, and associated overhead and
    trade-offs are discussed.

\end{abstract}

\begin{keyword}
Checkpoint-Restart \sep Fault-Tolerance \sep DMTCP \sep Infiniband \sep Multilevel Checkpointing, MPI Oversubscribing
\end{keyword}
\end{frontmatter}

\section{Introduction}

The trend towards parallel high-performance computing systems with extreme
numbers of cores, deep memory hierarchies, and multidimensional topological
networks is pushing application developers towards programming models that must
take advantage of nodes executing a large number of threads, while also
maintaining efficient internode communication. The evolution of
\textit{hybrid} programming models consequently results in parallel applications
that are effectively operating multiple runtime systems simultaneously to carry
out its computation. The common MPI+OpenMP approach is an example. In such a
context, it is reasonable to allow programming models to collaborate when
performing some runtime actions.

Consider the objective of \textit{checkpoint/restart} (C/R) as a fault-tolerance
mechanism aimed at saving the current state of a given parallel program's
execution (i.e., \textit{checkpoint}) and then restoring the program's status at
that point (i.e., \textit{restart}). There are multiple
methods to achieve this purpose:

\begin{itemize}
	\item  \textit{explicit} requiring direct modifications in the code;
	\item \textit{transparent} in the sense that they are able to checkpoint indifferently from the code itself.
\end{itemize}

One of the major stakes for end users is to select the C/R method fitting with
application needs. This paper presents  C/R optimizations leveraging runtime
support for both these cases. Transparently checkpointing complex applications
may lead to challenges when involving for example high-speed networks. The
application-level approach defers the C/R support to developers,  using their
knowledge to checkpoint only relevant data, whereas the runtime is, in most
cases, more suited to deal with low-level notions like C/R data replication. In
these two examples, we show how the runtime may collaborate to enable more
efficient checkpointing.

In the rest of this paper we consider application-level and transparent
checkpointing methodologies. We describe their respective implementation with
the \emph{Fault Tolerance library}\cite{FTI} (FTI) and \emph{distributed
multithreaded check-pointing} (DMTCP)\cite{ansel2009dmtcp}, contrasting their
use and purpose. Section \ref{related} starts by describing related work and
Section \ref{related:discussion} discusses various levels for
checkpoint-restart and tradeoffs. Then, Section \ref{sec:3-mpc} presents the
specificities of our MPI runtime executing MPI processes in user-level threads.
The rest of the paper eventually describes and validates the integration of two
fault-tolerance tools, FTI and DMTCP, with a focus on runtime oriented
optimizations.  More generally, we make the following contributions:

\begin{itemize}
	\item We show how application-level checkpointing could rely on dedicated progress threads, positively taking advantage of oversubscribing;
	\item We demonstrate high-speed network checkpointing thanks to collaboration from MPI runtime;
	\item We introduce a compact collective checkpointing call for transparent C/R.
\end{itemize}

This work is an extended version of a paper originally focused on the sole
integration of DMTCP in a thread-based MPI runtime\cite{MPCDMTCP}. This new
version features extended descriptions, additional contrast with respect to
application-level checkpointing and more generally C/R trade-offs. In addition,
we describe how we integrated FTI application-level checkpointing library  to
take advantage of user-level threads.

\section{Related Work}
\label{related}

Fault tolerance in the context of HPC applications is a very active field.  The
increasing complexities and constraints on parallel systems, combined with
falling \textit{mean time between failure (MTBF)} on systems with millions of
components, motivates the development of technology to mitigate the consequence
of failures during parallel execution.  Such failures directly map to lose
simulation results, but also the financial cost of a highly priced resource.
Beyond fault tolerance, these technologies can also benefit other purposes,
such as steering of a parallel application to improve solutions or remapping
system resources to address allocation constraints on a given machine.  With
respect to MPI applications in general, we can identify three main approaches
for fault tolerance: (1) explicit and (2) transparent approaches, followed by
(3) failure mitigation.  Although these are not mutually exclusive, we describe
each in turn.

\subsection{Explicit Methods}

The checkpoint/restart methodology is about both saving and restoring the state
of a program.  When it comes to parallel applications, this supposes that a
program (e.g., a simulation) is able to restore its state (data) and current
time-step (control) to take over the computation from where it was
checkpointed.  The most basic way to achieve this behavior is to manually save
data associated with a given time-step and reload it again to restart it, this
being done by the program itself.  In this manner, results from multiple
intermediate time-steps can be saved and reloaded.  This is a portable method
which has the advantage of not requiring any external tool.  The application
describes which data has to be saved and the resultant checkpoint file contains
exactly what is needed for restart while the program interruption time remains
low, keeping a small overhead for the overall application execution time.  One
step further is to consider checkpoint file storage in a redundant manner.  An
easy way is to store files on a shared mount point.  However, this approach
exposes issues when scaling to thousands of nodes/processes.
SCR\cite{moody2010design} and ACR\cite{ni2013acr} answer this by storing
checkpoint files over faster, local mount points and replicate them to ensure
redundancy. The \emph{Fault Tolerance library (FTI)}\cite{FTI} that we describe
in more detail in Section \ref{sec:2-fti} is also aimed at solving these
issues.

Unfortunately, the basic approach has further limitations.  First, it requires
that the full dataset remain easily serializable, and supposes that all the
artifacts linked to a given computation state are preserved and restorable.
This can be a difficult task when dealing with highly modular frameworks
hosting several data structures.  Second, it supposes that each simulation
implements its own checkpoint format and dedicate development efforts to
provide a similar feature.

As far as application-level checkpointing implementation mechanism is
concerned, incremental checkpointing~\cite{incremental} was proposed to reduce
the amount of data to write in consecutive checkpoints, but the benefits of
this technique are not always important.  Thus, disk-less
checkpointing~\cite{diskless} was proposed to alleviate this issue. With the
arrival of new storage devices, multilevel checkpointing was
proposed~\cite{FTI,multilevel}, including a certain number of features, such as
asynchronous transfers to the parallel file system.  Semi-blocking algorithms
have been proposed to save the checkpoint data without stopping the application
execution\cite{semiblocking}, however this work does not leverage threading
mechanisms as the one presented in this paper to safely and efficiently
oversubscribe compute nodes and allow fault tolerance tasks to take place in an
opportunistic fashion.

Oversubscribing, an approach we retained in this paper, has been scarcely
studied.  A complete survey of oversubscribing with the use of several parallel
programming languages~\cite{ihbz2010} shows that oversubscribing MPICH-2 MPI
processes induces an overhead of 10\% (equivalent to the one we observed with
OpenMPI), while oversubscribing threads may improve overlap and recovering
waiting periods. It has also been studied how bad placement of processes for
checkpoint/restart may hurt performance~\cite{wsr2015}.  Another work describes
how, even if possibly harmful inside one application, oversubscribing can be
used to efficiently execute multiple applications sharing one
node~\cite{Utrera2014}. To circumvent this drawback when applying MPI
oversubscribing in a unique application, some work focused on enabling multiple
MPI process in one OS process~\cite{kamal2012}, verifying the positive impact
of such implementation.

While it is possible to leverage external libraries that optimize certain
support, application-level checkpointing still requires representative data to
be manually described using a dedicated API.  As a consequence, they cannot be
seen as transparent, as the target code still has to insert calls to the
checkpointing API.  For these reasons, methodologies which do not involve such
annotations, have also been explored.

\subsection{Transparent Methods}

Transparent checkpointing tries to save the state of a running program, without
having any previous application knowledge.  Several tools have been developed
for this purpose, leveraging multiple approaches.  A general ``external''
method utilizes a virtual machine (VM) running inside an emulator, which can be
frozen and then saved (both from memory and disk point of
view)\cite{QEMU1,QEMU2}.  While effective, it requires the whole operating
system to be saved, and has severe performance overhead.

Tools for checkpoint/restart that are more appropriate for the HPC field
include the Berkeley Lab Checkpoint Restart\cite{BLCR} (BLCR) tool.  BLCR
relies on a kernel-level approach to both suspend and checkpoint.  This has the
advantage of avoiding a complete wrapping of every system call, and thus avoids
the associated overhead.  Being part of the Linux kernel also give the
advantage to restart applications in the exact same UNIX environment (same
process ID, restoring UNIX pipes).  However, the kernel approach first requires
an administrator to load the corresponding module.  Without considering
resources outside of the current OS, it is not possible to save/restore network
communication like sockets, and the application will have to handle these
limitations to provide a complete C/R support.  As multiple patched kernels are
not able to communicate through a whole cluster, BLCR, on its own, cannot be
used in MPI context.  The parallel application has to integrate explicit BLCR
support to enable its distributed usage.  In particular, an approach using BLCR
similar to what we present in this paper has been developed with the idea of
closing network resources before checkpointing~\cite{BUNTINAS200873}.  However,
it was limited to TCP protocol and considered emulation on high-speed networks.

Another approach consists in providing checkpointing in user-space by wrapping
any needed system calls, in order to constantly track application states.
Indeed, as tools cannot be sure about the application behavior, all potential
calls involving resources outside the process are to be captured, such as
network or storage.  The \emph{Distributed Multi-threaded CheckPointing}
(DMTCP)\cite{ansel2009dmtcp} tool can checkpoint applications at user-space
level, injecting a preloaded shared library upon application start in order to
wrap system calls.  Such a tool has the advantage of not requiring recent
kernel features or administrative privileges for installation or recompiling
the application to enable, disable or update the support.  From this viewpoint,
it becomes easier to make multiple nodes collaborate, and checkpointing
distributed applications does not necessarily need MPI-aware implementations.
However, catching system calls and associated bookkeeping creates a measurable
performance overhead.  Moreover, a log of on-the-wire messages has to be
preserved in order to replay them in case of a failure.  Such a model
introduces a non-negligible cost for the application.

The last method allows transparent checkpointing without wrapping system calls,
as done by tools such as CRIU \cite{CRIU}.  However, it relies on more recent
kernels to be able to fully extract information from the operating system.
CRIU has the advantage of supporting name-spaces and is, therefore, the
solution of choice when dealing with containers.

As far as MPI support is concerned, only DMTCP and BLCR currently integrate a
mechanism to enable a distributed checkpoint involving multiple UNIX processes.
For this reason, and due to test environment constraints (i.e., kernel), the
transparent solution we will develop in the rest of this paper relies on DMTCP,
but CRIU is recognized as a promising future alternative particularly as it
does not create additional overhead due to wrapping.

\subsection{MPI Failure Mitigation}

The failure mitigation approach is more focused on how to identify and put up
with a failure than actually on how to recover from it.  For example, if some
nodes suffer from a hardware failure during a MPI job, it would be faster for
the application to recover from remaining MPI processes than restarting the
whole program (reallocating resources)\cite{FTMPI, RTHR}.  If the workload can
be adjusted dynamically, such approaches are bound to be more efficient than
pure C/R.  In this field, we can cite the User-Level Failure
Mitigation(ULFM)\cite{bland2012evaluation, ULFM2}, a solution implemented on
top of OpenMPI, providing new MPI semantics that helps the application to
recover process failures.  This model defines a \emph{state} at the
communicator level.  If at least one MPI process becomes unreachable -- for any
reason defined by the implementation -- the MPI call returns an error.  In
addition, ULFM provides routines to revoke and shrink communicators in order to
recover from failures.  This approach can be made straightforward by attaching
an "error-handling" routine to the MPI interface, somehow analogous to signal
handlers on UNIX systems, they allow a given program to react appropriately to
a failure.  ULFM is therefore an MPI toolbox for resiliency in MPI context, and
should be seen as complementary approach to C/R.  One drawback of such
interface is that it is still up to the application to implement the part of
the code dedicated to failure mitigation\cite{PLANB,LFM, FENIX}.

\subsection{Summarizing Checkpointing Approaches}
\label{related:discussion}

From a general point of view, transparent methodologies have the drawback of
saving more than needed for a given execution.  Indeed, it is not compulsory to
save internal runtime states to restore a given simulation.  Nonetheless, in
some cases, it may not be sufficient to solely rely on data restore.  For
example, a given computation may use data types which are solely created during
program startup.  As a consequence, a program based on application-level
checkpointing also has to go through an initialization phase of some form, to
restore pertinent resources. There is then a clear tradeoff between these
approaches.

\begin{table*}
	\centering
	\begin{small}
		\begin{tabular}{|c|c|c|c|c|c|c|}
			\hline
			Checkpoint Level & Selectivity & Size & Administrative Rights & Implementation. Cost & Transparent & Overhead \\
			\hline
			\hline
			Application & {\color{ForestGreen}High} & {\color{ForestGreen}Small} & {\color{ForestGreen}No}  & {\color{red}High} & {\color{red}No} & {\color{cyan}Variable} \\
			\hline
			User-level & {\color{orange}Low} & {\color{orange}Large} & {\color{ForestGreen}No} & {\color{orange}Small} & {\color{orange}Mostly} & {\color{orange}Medium} \\
			\hline
			OS-level & {\color{red}Lower} & {\color{red}Larger} & {\color{orange}Yes (install)} & {\color{ForestGreen}Null} & {\color{ForestGreen}Yes} & {\color{ForestGreen}Low} \\
			\hline
			Hypervisor & {\color{red}Lower} & {\color{red}Larger} & {\color{orange}Run virtualized} & {\color{ForestGreen}Null} & {\color{ForestGreen}Yes} & {\color{red}High}\\
			\hline
		\end{tabular}
	\end{small}
	\caption{Comparison of the various checkpoint/restart levels}
	\label{crcomp}
\end{table*}

As presented in Table \ref{crcomp}, we compared various levels for checkpoint
restart and describe the tradeoff they incur. The levels we considered, can be
described as follows:

\begin{itemize}
	\item \textbf{Application-level}: adding code or using dedicated libraries to implement checkpoint-restart inside the target application, for example using FTI;
	\item \textbf{User-level}: implementing (transparent) checkpoint restart through either state capturing or system call wrapping in user-space, for example with DMTCP;
	\item \textbf{OS-level}: (transparent) checkpoint/restart thanks to Operating System (OS) support possibly through a dedicated kernel module, for example with BLCR;
	\item \textbf{Hypervisor}: using capabilities of the virtualization environment to suspend, save and restart (transparently) a running virtual machine, for example with QEMU\cite{QEMU1}.
\end{itemize}

It can be seen that no solution is ideal, indeed being transparent comes at the
cost of more overhead due to system-call wrapping and bookkeeping. Moreover,
transparent methodologies are not able to extract the minimal dataset linked
with application state and control, and generally save the complete memory
image (either application or sometimes the full OS). This leads to larger
checkpoints offset by benefits in development time as no modifications to the
code are needed. The right checkpointing method is then probably a mix of those
presented in this table and is yet to be defined. In this , we decide to not
focus on one specific method but instead proposing multiple solutions on top of
your MPI implementation, offering the user the ability to choose the right one
for his own scenario. In the next section we are going to describe the
thread-based MPI runtime in which we propose to integrate transparent and
application level checkpointing, providing an  initial context to the later
developments.

\section{MPC Overview}
\label{sec:3-mpc}

MPC~\cite{pjn2008} is a framework dedicated to the smooth integration of
shared-memory parallel programming models in MPI applications.  To this end,
MPC provides different implementations such as MPI, OpenMP, and Pthread, all
unified on top of the same user-level thread scheduler.  By having its own MPI
implementation and its own thread scheduler, MPC is then able to execute MPI
processes in different configurations, as discussed below. One can consider
that all MPI implementations fit into one of two categories: process-based and
thread-based.

\emph{Process-based implementations} are based on MPI Processes being regular
UNIX processes, with separate address spaces. Most MPI implementations fit in
this category, such as MPICH and OpenMPI. An indirect consequence of this state
of things is that applications may feature global variables duplicated for each
MPI process running as a UNIX process.

\begin{figure*}
	\centering
	\subfigure[Process-based model]{
		\includegraphics[width=0.45\textwidth]{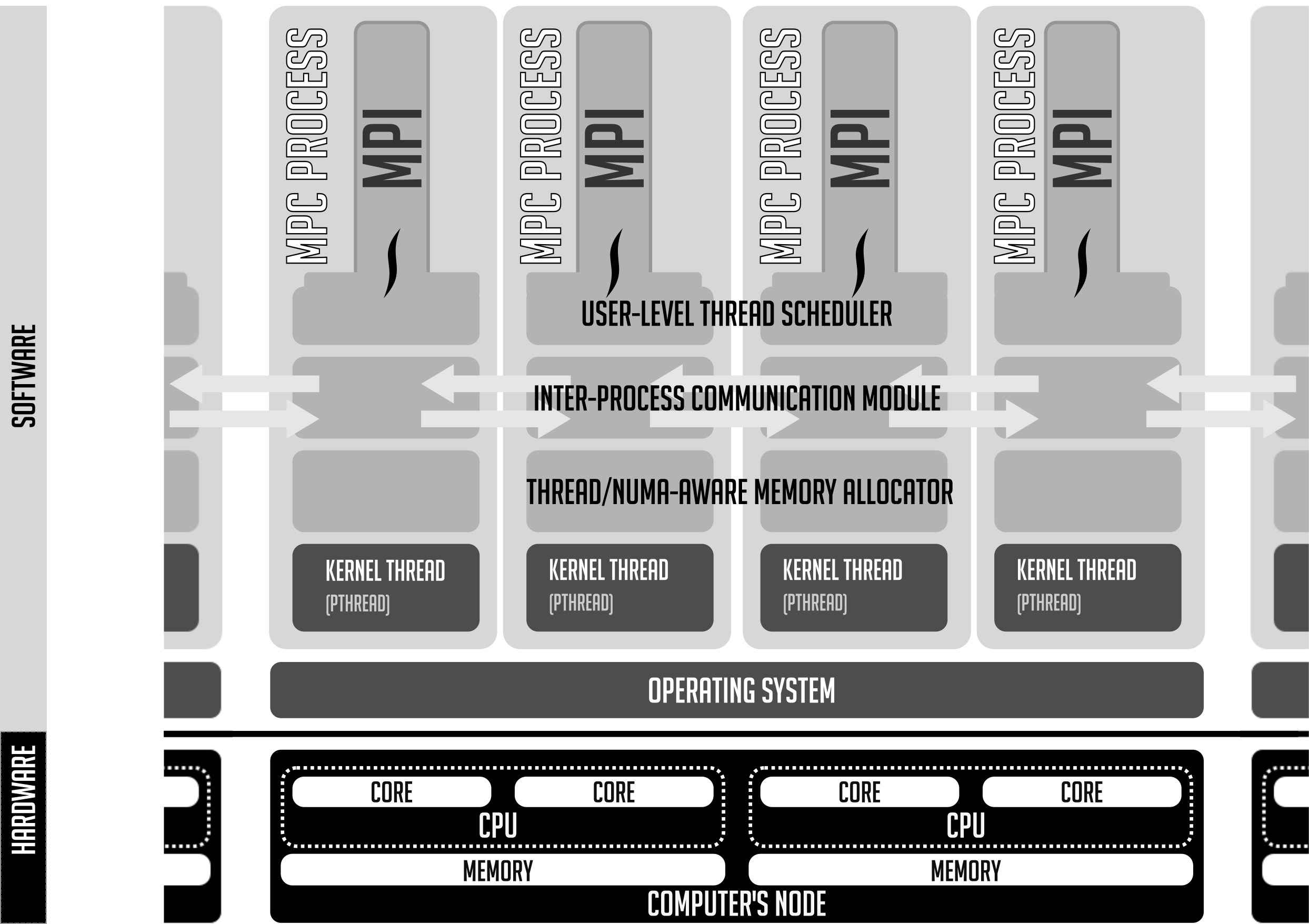}
	}
	\subfigure[Thread-based model]{
		\includegraphics[width=0.45\textwidth]{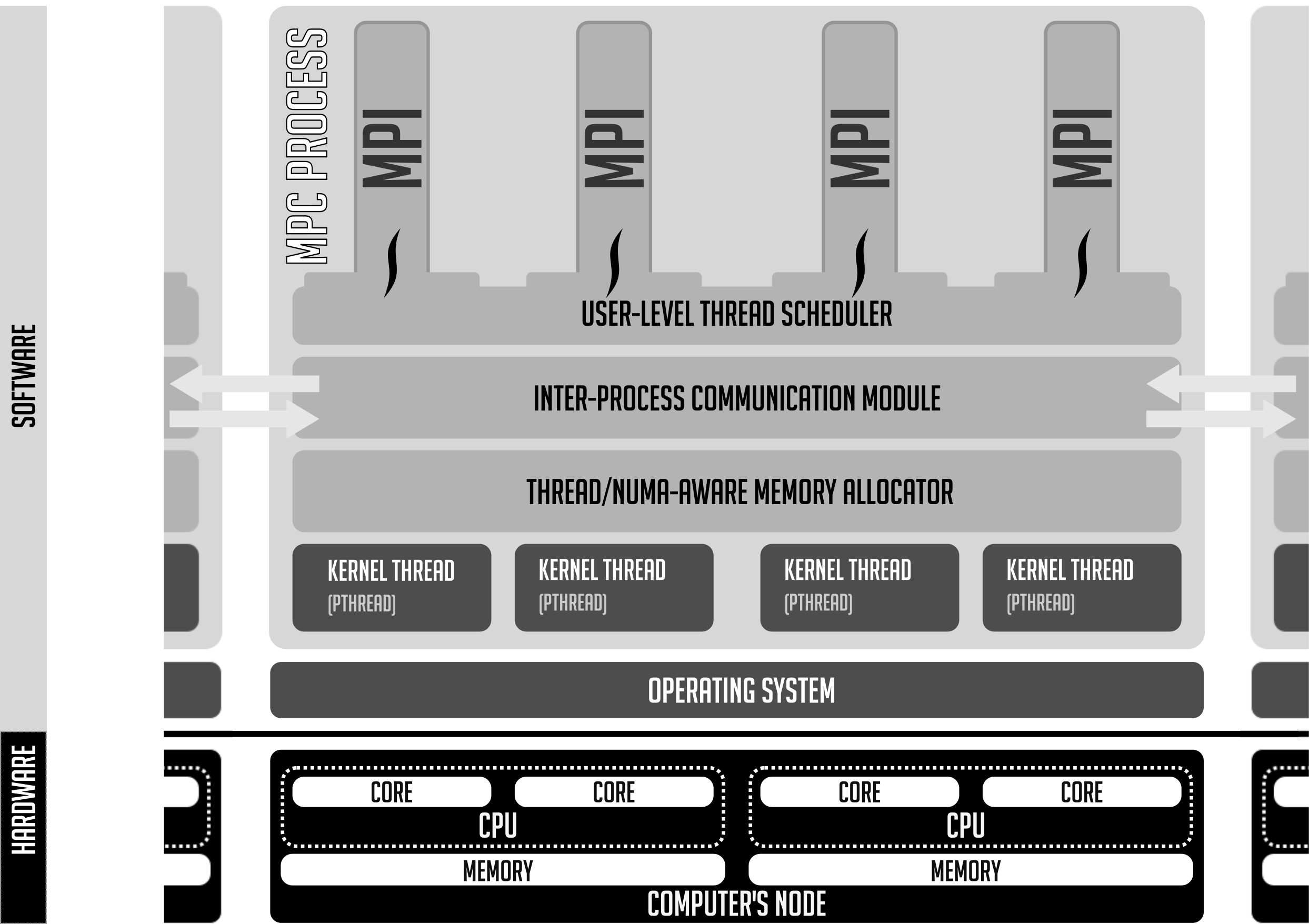}
	}
	\caption{\label{fig:mpi_flavors} MPI implementation flavors}
\end{figure*}

\emph{Thread-based implementations} are using threads for running MPI
processes. In this second configuration, multiple MPI Processes run in a single
UNIX process and share memory. A hybridization is also possible by running
multiple UNIX processes, themselves gathering several MPI processes. In this
context, global variables are not fully isolated anymore, by construction.

%\newline

To address this second configuration, MPC relies on a privatizing compiler to
transparently separate global variable by creating multiple copies of it for
each MPI process, thanks to a hierarchical TLS storage
approach\cite{besnard2016introducing}. It is then possible to port C, C++ and
Fortran applications with little effort to this thread-based configuration. The
advantage of running in such configuration is that the scheduling of MPI
processes does not rely necessarily on the OS scheduler, but on a user-level
thread scheduler. In addition, intranode communication is simplified, in that
regular shared memory copies can be used, instead of relying on SHMEM or CMA to
achieve the same effect. MPC encompasses these two  ``flavors'' of MPI
processes\cite{pcj2009}. These configurations are displayed in
Figure~\ref{fig:mpi_flavors}.  On the left, the usual process-based MPI model
completely separates components for each MPI process. With the process-based
flavor, each MPI process has its own thread scheduler instance, its own
allocator instance, working on independent address spaces. On the right, thanks
to the thread-based approach, more components are shared between MPI processes.
This induces a reduction of the global memory footprint since some internal
structures (such as network buffers) are not duplicated.  Besides, multiple MPI
processes on the same node rely on the same MPC scheduler, potentially
bypassing the OS scheduler

\section{Contribution}
\label{contribution}

Here we explore more extensively two use cases of checkpoint restart in the
context of a thread-based MPI runtime. First, we consider the integration of
DMTCP in the MPC runtime to provide transparent checkpoint-restart
capabilities. Second, we focus ourselves on the FTI library and how it was
integrated in MPC's unified scheduler. For each of these examples we will
follow the same plan. We will introduce the approach in general, contrasting it
with other methodologies and explaining how it is implemented. Then, we will
detail how it was integrated in our thread-based runtime. Eventually, we
present for each model a commented performance measurement, outlining
respective advantages and limitations.

As far as transparent checkpointing is concerned, we present a simple
collective call enabling checkpointing at coherency points with respect to
communications. In addition, we detail how we developed a signaling network
enabling transparent restart through DMTCP. Our performance results on Lulesh
at scale, show that transparent checkpointing is possible without sacrificing
network performance. However,there is a trade-off with respect to checkpoints
as our methodology supposes that the high-speed network is closed during each
checkpoint, requiring a transitive overhead which can still be mitigated as we
further elaborate.

Dealing with application-level checkpointing, we considered the FTI
checkpointing library and explored alternative approaches with respect to the
mapping of helper processes. Indeed, by default, FTI relies on MPMD to map a
process in charge of data-replication on each node. In our model, we moved this
process inside a thread running in an oversubscribed fashion over an user-level
scheduler. In particular, we show slight improvements thanks to this
collocation in threads, instead of processes. But still, as MPC does not
integrate scheduling points inside I/Os the overall gain remains limited in
terms of I/O recovering.

Overall the rest of this paper is an assessment of how checkpointing could be
integrated inside a thread-based MPI. We show not only that it is possible with
some advantages. But the most important conclusion is that doing so is not
different from what would be done with a "regular" process-based MPI. In fact,
for the transparent C/R examples concerns were focused on how to save and
restore the network state aspect which is directly translatable to other MPIs.

\section{Transparent Checkpoint-Restart with DMTCP}

In this Section we focus ourselves on transparent checkpointing inside the MPC
thread-based MPI runtime. The goal of such approach is to enable C/R with
limited developments (hence the transparent adjective). However, to achieve
such thing, a careful handling of application's state is compulsory. Indeed,
while checkpointing a serial program is straightforward, the distributed nature
of an MPI application requires a restoration of the network connectivity. The
rest of this section aims at describing how we made such restart possible in
MPC MPI. To do so, we first introduce DMTCP, then after recalling MPC's network
structure (bootstraping, multi-rail) we cover the integration of DMTCP itself.
To do so, we present a simple collective checkpointing interface that dodges
the complex issue of in-flight messages. Eventually, we conclude this section
with a performance study done with the Lulesh benchmark.

\subsection{DMTCP Overview}
\label{descdmtcp}

We consider checkpointing through the user-level transparent approach using
DMTCP~\cite{ansel2009dmtcp} which is the distributed implementation of
MTCP~\cite{rieker2006transparent}, a user-level checkpoint implementation
compatible with POSIX threads.  Its goal is to transparently save and restore
distributed applications.  To do so, it relies on a coordinator process
(\texttt{dmtcp\_coordinator}), steering applications under C/R for the current
user.  It can be reached through an IP address/port tuple.  Users can then
interact with the coordinator through running applications or the CLI.  Each
application to track is wrapped with the \texttt{dmtcp\_launch} command,
preloading the MTCP wrapping library, on each process to start.  By wrapping
most of the libc, DMTCP is able to closely track the relationship between
execution streams.  Moreover, a signal handler is defined in each thread (by
default \texttt{SIGUSR2}), to trigger a checkpoint, stopping each thread (using
\texttt{tkill}), saving its own data, including local context (register) and
stack.

At the network level, DMTCP is able to save alive sockets and pipes (after
converting them to socket pairs).  For this purpose, it goes through a
comprehensive process including the election of an owner of the respective file
descriptors (when shared between forks) and accounting for "on wire" data
inside the socket in order to restore them in case of a restart.  As a
consequence, DMTCP can reliably save TCP connections between distributed
processes in a transparent manner.  It is this aspect we rely upon for MPC.
Also, DMTCP is able to save shared-memory segments, making it compatible with
processes running on the same node with SHM.

As far as the restart model is concerned, the first step is to recreate the
same topology, relaunching each checkpointed process.  DMTCP proposes a
dedicated script only compatible with Hydra and Slurm, ensuring the new
configuration (from the restarting environment) is compliant with the initial
one, before restarting the processes.  The first step deals with restoring
network connections (and pipes) as they might be shared between processes.
Then, execution streams are restored and eventually the program image is
injected from the checkpoint data and file descriptors are reopened and offsets
restored.  At this point, execution streams wait in a semaphore and are able to
restart once all threads are ready.  DMTCP reproduces the same process and
thread hierarchy (by tracking fork/clone) to make the system topology
consistent (parent/child relation).

It is this process, fully accounted for by DMTCP, that we leverage in MPC to
provide the checkpoint-restart  feature with the subtlety of hosting several
MPI processes in a single UNIX process.  In this case and as we will further
describe, a dedicated synchronization mechanism is required.

\subsection{Network Modularity in MPC}
\label{modular}

As our support for transparent checkpointing is based on the ability to close a
network rail and then restoring it, this section is dedicated to describing how
we managed to expose sufficient modularity in our communication layer to enable
such support. MPC's low-level architecture is based on \textit{communication
rails} which are associated with a given network driver.  MPC can combine at
runtime multiple communication drivers, which are used together to provide
communication capabilities at the MPI interface level.  In this section, we
present an overview of ``multi-rail'' support in MPC.  In particular, we
discuss how MPC is able to bootstrap its network using \emph{control messages}
(``signaling'' messages) routed on a base topology.  With this mechanism in
place, we outline how it contributes to transparent checkpointing by preserving
high-speed network capabilities.

\subsubsection{Multi-Rail}

MPI is dedicated to enabling high-performance messaging between distributed
processes.  To do so, it can rely on multiple network technologies. For
example, one system could use Infiniband EDR between nodes and a shared-memory
segment (SHM) inside a given node at the same time.  More generally, an MPI
runtime usually supports at least two network types: (1) for optimized
intra-node communications (latency lower than the $\mu$sec), and (2) for
internode communications, where remote-direct memory access (RDMA) support
could be used to optimize MPI's performance (in the $\mu$sec range).  The
switch between intranode and internode policies is then defined as the position
of the target MPI process relative to the source, that is, whether they are
located on the same node.  Multi-rail is then naturally present in any
state-of-the-art MPI runtime.  The following describes how MPC handles
multi-rail, but the overall principle is applicable to any MPI runtime,
thread-based or not.

As shown in Figure \ref{fig:mpi_flavors}, MPC is a thread-based MPI
implementation which makes it possible to have multiple MPI ``tasks'' within a
MPI ``process'' that is bound and running in a UNIX process.  MPI tasks are
equivalent to traditional MPI processes in that they can communicate via MPI
with each other.  Thus, message headers in MPC carry both MPI process id
(internal to MPC) and task id.  The process id is used to determine which UNIX
process hosts a given MPI process.  As a consequence, there is no direct
correlation between a communication endpoint and a given MPI process.  Indeed,
a given network is only initialized once per UNIX process and therefore
multiple MPI processes will share the same network layer. Several situations
can result.  For instance, messages could be exchanged inside a given MPI
``process'' if both tasks are running in shared memory.  Or the messages could
be routed to the multi-rail network layer if the tasks are remote from each
other.  In this case, the multi-rail support must identify the most efficient
rail to reach a given remote UNIX process.  Moreover, these means of exchange
are not mutually exclusive and hybrid configurations involve both messaging
layers depending on peers.

In the rest of this paper, as far as transparent checkpointing is concerned, we
will consider solely the communication between UNIX processes, as it is the
only part of MPC involving internode communications interacting with network
cards. This put us in the process-based case where MPI processes are UNIX
processes and allow us to reason in a more general context applicable to any
MPI implementation. However, it should be noted that the methodology we develop
in this paper has been validated in all configurations of Figure
\ref{fig:mpi_flavors}.

\begin{figure}
	\centering \includegraphics[width=0.8\linewidth]{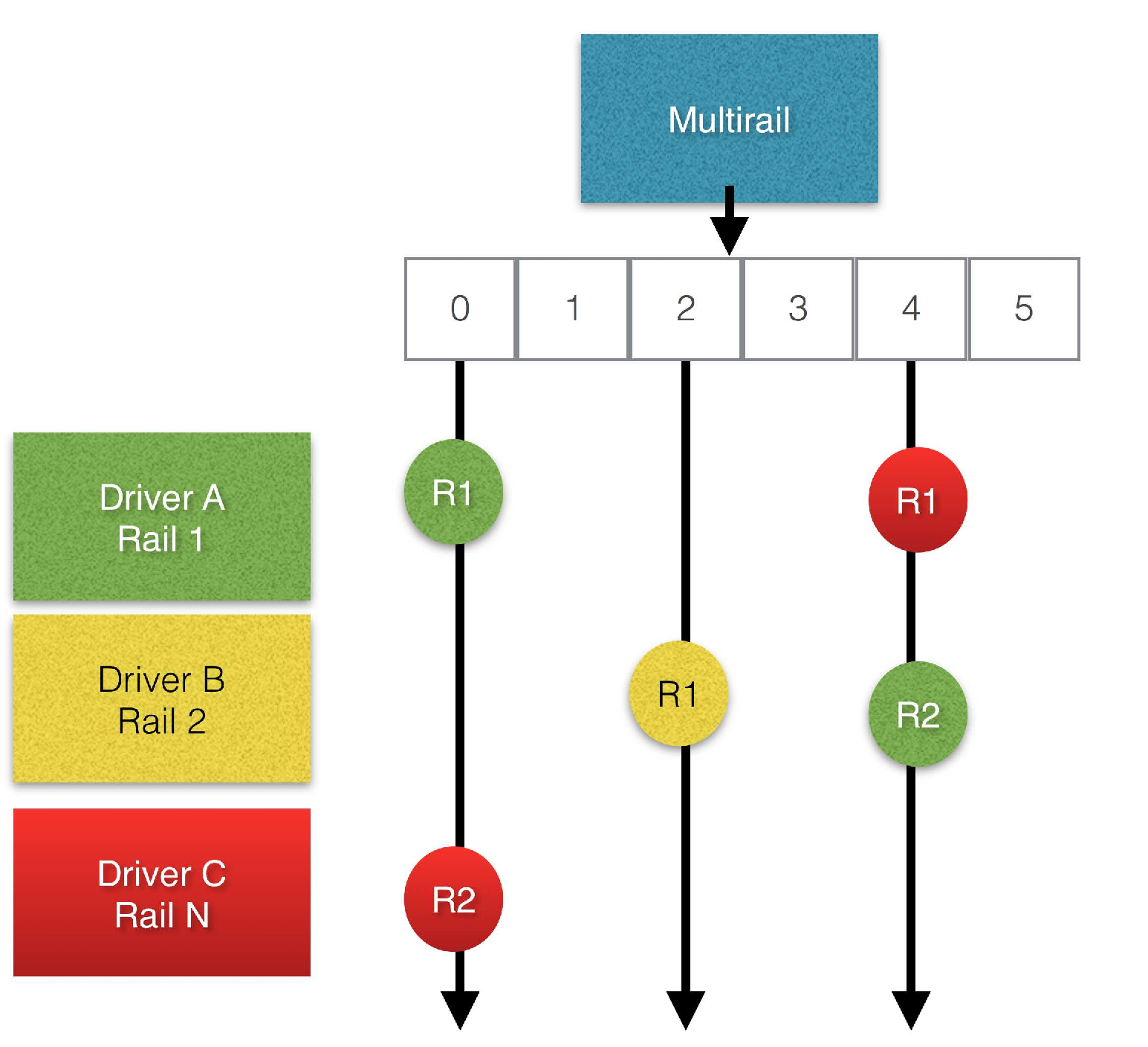}
	\caption{Overview of the multi-rail infrastructure in MPC.}
	\label{fig:mr}
\end{figure}

Communications in MPC are based on endpoints belonging to a communication rail
(see Figure \ref{fig:mr}).  Endpoints are sorted by priority inside ordered
lists corresponding to a given remote process.  When MPC tries to communicate
with a remote endpoint, it walks on the list for a given endpoint and tries to
\emph{elect} a candidate.  Election includes the concept of \emph{gate},
setting conditions (in terms of message type or size) related to the use of a
given rail.  If no endpoint is found in the list, a second election process is
done walking rails in order of priority to create a new endpoint.  If one rail
supports this \emph{on-demand} feature, the connection handler is called in
order to create the low-level route.  Section \ref{bootstrap} will detail how
this \emph{on-demand} connection process is implemented in MPC.  If this
succeeds, MPC proceeds to use the new endpoint.  Otherwise, it crashes with a
\emph{no route to process} error, meaning that no valid network path exists or
could be created to reach the targeted process.

\begin{figure}[ht!]
\begin{lstlisting}[language=XML]
<config>
	<name>tcp_config_mpi</name>
	<driver><tcp/></driver>
</config>

<rail>
	<name>tcp_mpi</name>
	<priority>1</priority>
	<topology>ring</topology>
	<config>tcp_config_mpi</config>
</rail>

<rail>
	<name>tcp_large</name>
	<priority>10</priority>
	<topology>none</topology>
	<config>tcp_config_mpi</config>
	<gates>
		<gate>
			<minsize>
				<value>32KB</value>
			</minsize>
		</gate>
	</gates>
</rail>

<cli_option>
	<name>multirail_tcp</name>
	<rails>
		<rail>tcp_large</rail>
		<rail>tcp_mpi</rail>
	</rails>
</cli_option>
\end{lstlisting}
\caption{Example of XML configuration file for MPC's multi-rail engine.}
\label{fig:xmlconf}
\end{figure}

As presented in Figure \ref{fig:xmlconf}, MPC's multi-rail support relies on an
XML configuration file that we now describe bottom to top.  First, we define a
Command Line Interface (CLI) option named \emph{multirail\_tcp} and attach two
rail definitions to it: \emph{tcp\_large} and \emph{tcp\_mpi}.  As a
consequence, when launching the parallel execution with \emph{mpirun}, the
\texttt{-net=multirail\_tcp} option will create the two aforementioned rails.
If we now look closer at the rail definitions, each of them is named and is
attached to a priority.  Observe how the \emph{tcp\_large} has a higher
priority than the \emph{tcp\_mpi} one, it is because we want each message to
first try it.  Indeed, the "large" rail has a gate function defined and
requires a message to be larger than 32Kb to be able to transit through it.  If
this test fails, the message then checks the \textit{tcp\_mpi} rail which
matches any message (as it has no gate function).  One last part involved at
the beginning of the configuration file is the network-level parameters in the
\emph{config} markup.  In this case, they are shared between the two rails and
we simply use the default TCP configuration -- the end user is free to create
configurations for his own rails.

One point that we overlooked in the previous configuration is rail
\emph{topology}. It plays an important role in the checkpoint-restart mechanism
because it defines the initial connection state of rails (defined as
\emph{static} routes).  Such initial routes are used to convey \emph{control
messages}, allowing on-demand connection mechanisms to establish additional
networking configuration.  We refer to this as the \textit{signaling network}
for MPC.

\subsubsection{Signaling Network}

MPC's network layer can also be used to provide a signaling network whose role
it is to allow remote processes to be reachable in a one-sided fashion. This
could also be described as remote procedure calls (RPCs) or as active messages
(AM) in MPI semantics. Indeed, some MPI functionalities already depend on this
being possible, for example, when establishing \emph{on-demand} connection,
emulating one-sided when no RDMA capable network is available, and even within
the rendezvous protocol, where target notification is required.

One of the main proprieties of the signaling network is its capability to route
messages according to a simple 1D distance metric, defined as the absolute
value between the source and target ranks.  The reason for retaining such a
simple metric is because we wanted it to be portable on any topology,
indifferently from its complexity.  To do so, we imposed the simple constraint
of embedding at minimum a ring in the topology.  This ring is what we call
``static routes'' in MPC's bootstrap and its main role is to ensure that, for
given a source process, there will always be a path to minimize the 1D distance
to its destination.  It is this property that incited us to rely on a minimal
ring, since dealing otherwise with sparse and/or arbitrary topologies, there
may be cases where the 1D distance metric is not sufficient to escape from a
local minimum.

\begin{figure}
	\centering
	\subfigure[2D-Mesh]{
		\includegraphics[width=0.8\linewidth]{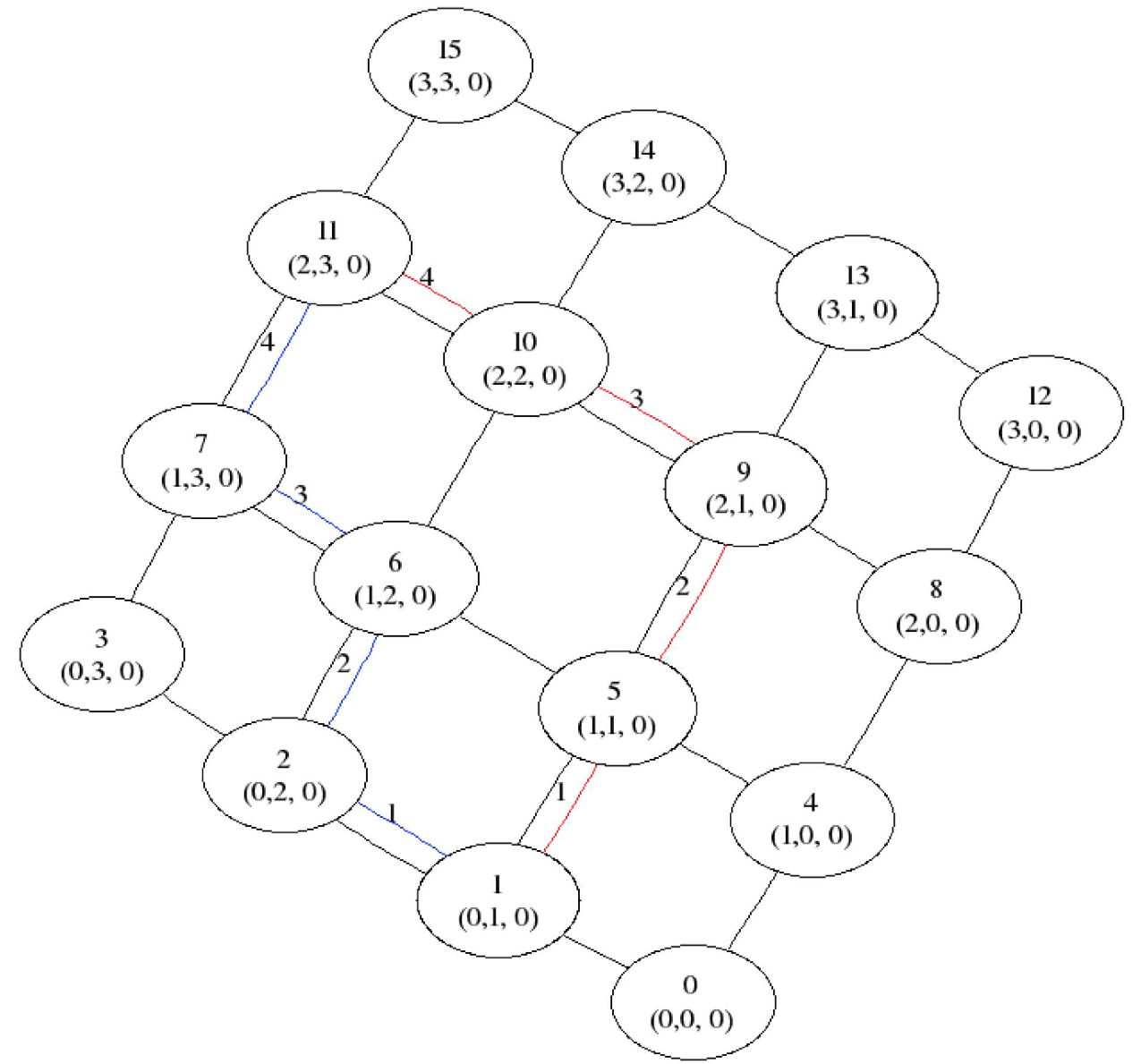}
	}
	\subfigure[3D-Mesh]{
		\includegraphics[width=0.8\linewidth]{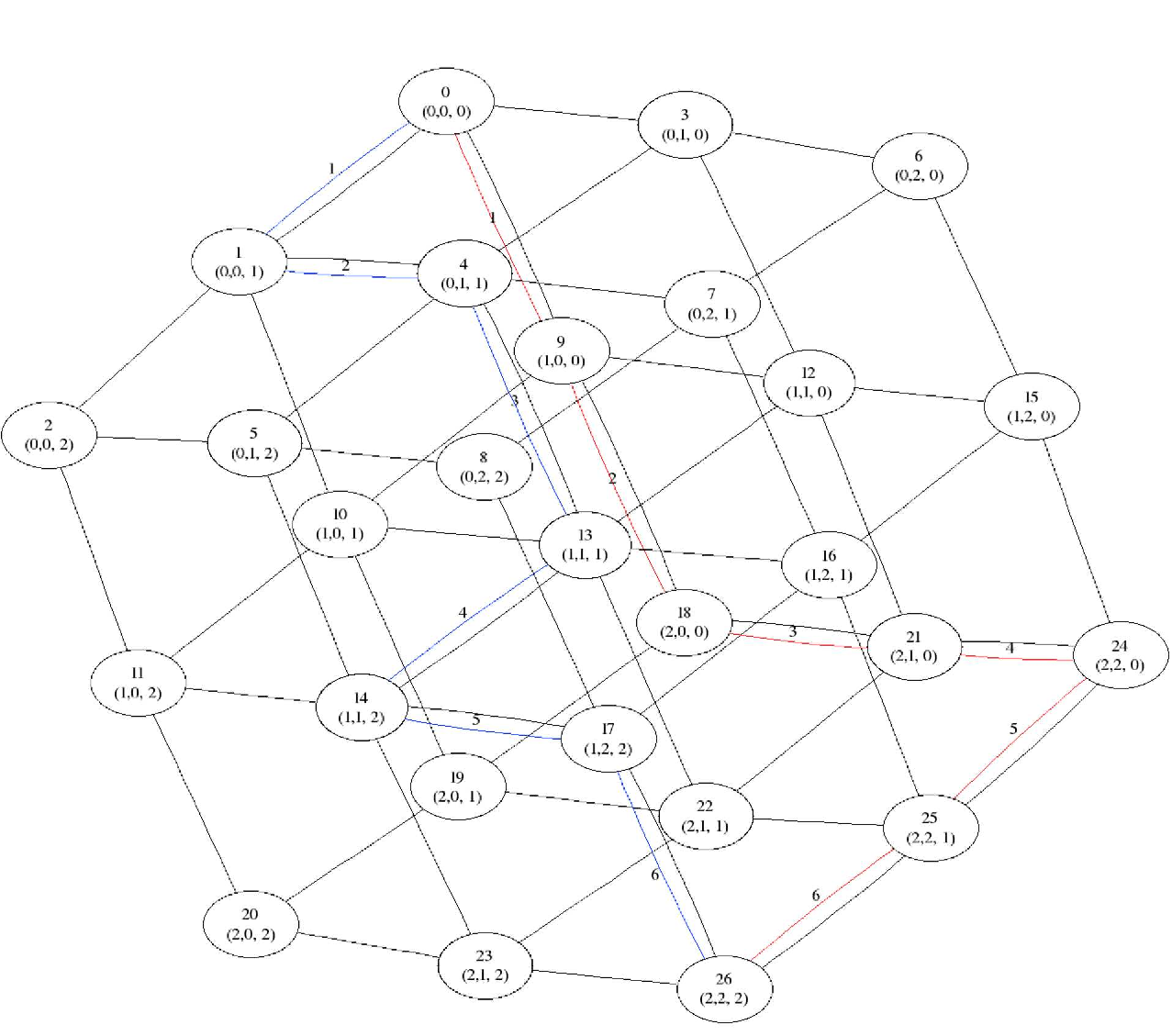}
	}
	\subfigure[2D-Torus]{
		\includegraphics[width=0.8\linewidth]{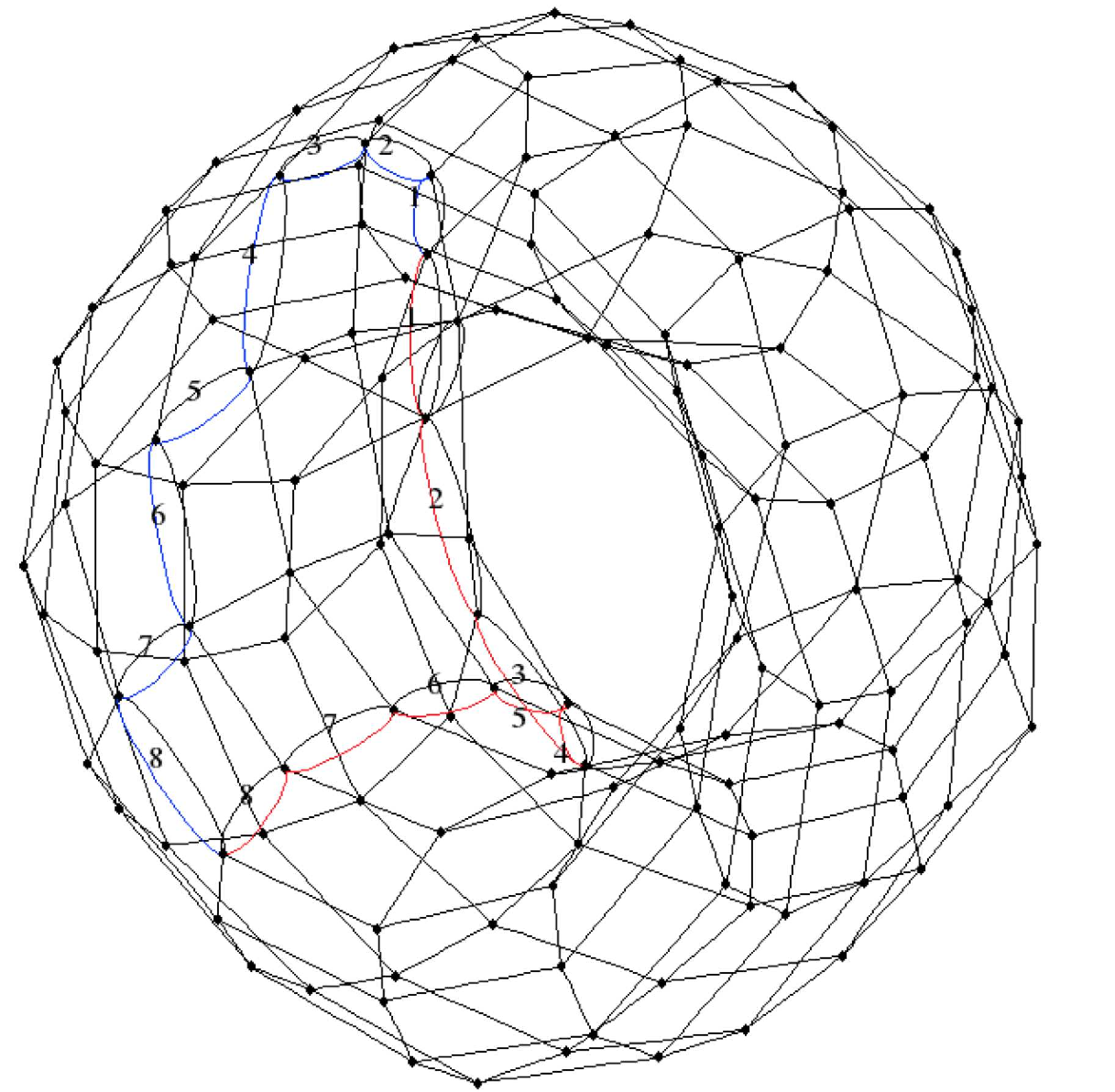}
	}
	\caption{
		Routing comparisons on various topologies with 1D in {\color{red} red}, and
		3D distances in {\color{blue} blue}.
		\label{fig:route3D}
	}
\end{figure}

Despite managing to limit the theoretical network diameter with denser
topologies, the 1D distance metric we retained to ensure routing robustness for
arbitrary topologies fails inidentifying the shortest path in higher
dimensions. Nonetheless, the availability of \emph{shortcuts} still allows 1D
routing to take advantage of the higher dimensions -- see for example the
difference of behavior between 1D and nD distances in Figure~\ref{fig:route3D}.

\subsubsection{Network Bootstrap} \label{bootstrap}

As presented in previous Sections, MPC implements a multi-rail engine and a
signaling network.  However, in order to be able to create endpoints, there are
some cases when a process would like to query information from another one
without knowing it explicitly.  The case which is of particular interest in
this paper is the on-demand connection when, for example, two processes are
exchanging and building their Queue-Pair information.  This can be illustrated
with a TCP analogy, where the IP address and remote port have to be exchanged
prior to establishing a connection.  When MPI starts, processes are usually
disconnected and connected on demand.  To do so, MPI runtimes rely on the
\textit{process management interface} (PMI) provided by the launcher.  PMI
provides a Key-Value Storage (KVS) which is relied upon to bootstrap
connections, prior to MPI processes creation.  MPC naturally relies on the PMI,
but it also implements its own bootstrap system in order to limit the amount of
information to be exchanged with the PMI.  Indeed, if there are thousands or
even millions of processes it can be costly to exchange the whole information
relative to all the ranks in an all-to-all manner, particularly prior to having
any high-performance communication substrate.  To circumvent this, MPC defines
the notion of rail topology.  In all cases, there must be a rail accepting all
messages with a ring topology (see \emph{tcp\_mpi} in Figure
\ref{fig:xmlconf}).  This rail is initialized using solely the PMI KVS,
exchanging, in this case, \emph{rank:host:port} tuples.

Later \emph{on-demand} connections, however, will not rely on the PMI, but on
control messages which can be routed through the network until their
destination.  Such messages use a distance metric and take advantage of any
route and any rail.  Consequently, even if only a TCP ring is present during
startup, it is highly probable that ``shortcuts'' will appear as MPI processes
start communicating.  This property is at the core of MPC's ability to
checkpoint-restart.  Indeed, existing checkpointing tools are not able to save
the network state for high-speed networks unless by wrapping all existing API
calls.  This leads to important overheads, for example, in the case of
Infiniband. Instead, such tools are limited to solely restoring TCP sockets
between processes.  As we will further discuss, this capability in MPC allows
restored MPI programs to operate immediately after the restart, instead of
relying on a complete network re-initialization through a PMI key exchange.

The main points to remember are MPC's multi-rail engine and its ability to
manage endpoints of multiple types to enable communication. These endpoints are
stored in an ordered list and go through an election mechanism.  MPC relies on
the PMI only to bootstrap an initial ring which is relied upon to convey later
on-demand connection requests. Thanks to its modular definition, MPC is capable
of closing a given rail removing all references to the associated network.  It
is this mechanism, combined with signaling, that enables MPC's transparent
checkpointing capabilities without wrapping network calls.

\subsection{DMTCP Support}

We leverage the DMTCP checkpointing tool to transparently save the state of an
MPI program.  In particular, we show how the MPI runtime can work with a
transparent checkpointing tool to enable support for high-speed networks. When
using specialized networking hardware, such as Infiniband (IB), care must be
taken with respect to initialization and handling of dedicated objects like
queue pairs.  Moreover, even if part of this context is saved in a transparent
checkpoint, restarting must avoid errors that could occur by launching the
program without setting up a connection to the Host-Channel Adapter (HCA)
within the process first. If we omit the question of high-speed networking,
checkpointing with DMTCP is transparent and relies on submitting requests to a
daemon in charge of the process without synchronization from the application or
the runtime. However, to deal with high-speed networks in a more efficient
manner, a contribution from the runtime is necessary to avoid large overheads.
We propose to leverage a dedicated modular network management infrastructure
developed in the MPI runtime to both reset and initialize networks on the fly
to enable such checkpoints.  As far as transparent checkpointing is concerned
this paper makes the following contributions:

\begin{itemize}
	\item
	The definition of a collective checkpoint interface enabling transparent
	checkpointing in MPI runtimes (Section \ref{iface});

	\item
	The concept of an in-band signaling network with the associated routing,
	and the use of multi-rail logic to enable partial checkpointing (Section
	\ref{modular});

	\item
	General MPI implementation of transparent checkpointing including
	high-speed networks.
\end{itemize}

This work has been implemented in the MPC thread-based MPI runtime, although it
is applicable to any MPI implementation, as we will describe.  What makes MPC
particularly challenging is that we needed to manage transparently the
checkpointing of multiple \emph{runtime stacking} configurations that MPC
supports.  Indeed, because MPC is built on user-level threading system, not
only does our approach track process-based MPI, but it can accommodate any type
of thread-based MPI, including user-level threads in MPC.  Thus, this
demonstration in MPC gives us confidence that the methodology will translate
well to future evolutions of MPI, including those supporting the concepts of
endpoints~\cite{dinan2014enabling} and sessions~\cite{holmes2016mpi}, which
involve intra-process parallelism.

Given the MPC infrastructure, the following presents a general methodology
enabling transparent checkpoint/restart for programs using high-speed networks.
More precisely, we detail how the MPC runtime is able to dynamically open and
close communication rails through a two-level checkpoint infrastructure.  Such
an approach provides MPI runtime with the ability to be checkpointed, and
transitively applications to benefit from this feature.  Moreover, we show that
this approach incurs a reduced performance overhead.

\subsubsection{Thread-Based MPI Checkpoint}

DMTCP and its coordinator are designed so that a single request for the
checkpoint is automatically broadcast to all the processes.  However, in MPC we
have to handle the fact that there are multiple MPI processes in a given UNIX
process -- checkpointing taking place at this latter level.

\begin{figure*}[ht!]
	\centering
	\includegraphics[width=.7\linewidth]{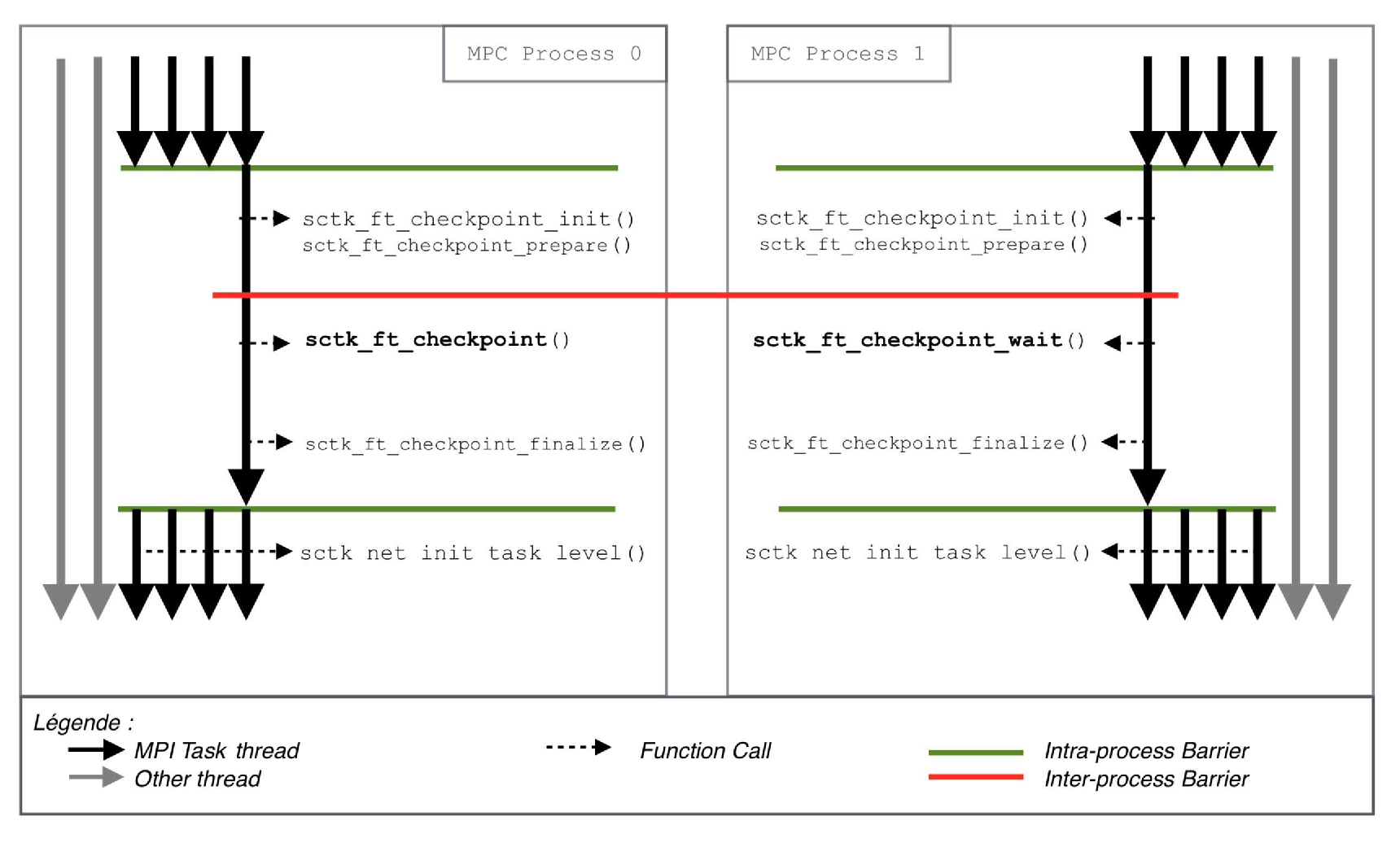}
	\caption{Two-level synchronization scheme enabling checkpointing in MPC.}
	\label{fig:thrs}
\end{figure*}

As depicted in Figure \ref{fig:thrs}, MPC solves this by proceeding to a first
intra-node barrier between MPI processes located in the same process.  Once a
master task has been elected, a second barrier occurs between processes such as
only a single rank invokes the internal checkpointing routines of DMTCP.

\subsubsection{Limitations in DMTCP}

During our developments around this integration of DMTCP in MPC, we discovered
limitations in the tool.  The developers have been very active to address some
of them and some others are still pending.  We will now provide a quick outline
for each of them.

\paragraph{Pinning Preservation} When we began our developments, pinning was
preserved at the checkpoint, but not at the restart.  The consequence was that
threads were not bound to a particular core.  This may remain unnoticed in the
case of a process-based MPI.  However, as MPC launches a single process per
node, this led to performance loss.  This has been reported and fixed in the
Git repository, in the branch tracking version 2.5.

\paragraph{Memory Locality} When a process is restarted with DMTCP, pages are
generally not located on the correct numa-node.  This leads to a loss of
locality for the restarted process.  To date, this has not been addressed in
DMTCP.  A possible workaround to this would be to rely on external tools such
as \textit{autonuma}\cite{corbet2012autonuma}.

\paragraph{GS Register Handling} In its current version, DMTCP does not save
the GS register.  This is generally harmless as this register is mostly unused
on x86\_64.  However, MPC uses this register to infer its own level of TLS
(Thread-Local Storage) indirection\cite{besnard2016introducing}, similar to how
\texttt{FS} register is used in common Pthread implementations for the same
reason.  As a consequence, an unmodified version of DMTCP is not able to
correctly checkpoint a privatized program (i.e., multiple MPI processes inside
a single UNIX process). This has been discussed with the
developers\footnote{\url{https://github.com/dmtcp/dmtcp/issues/607}} and we
proposed a fix.

\paragraph{Runtime defining pthread\_create} As MPC provides its own user-level
thread scheduler, it provides its own pthread implementation.  When being
wrapped by DMTCP, we encounter an issue as it is preloaded and implements
\texttt{dlsym}, yielding the following call stack:

\begin{lstlisting}
#0 pthread_create (from libdmtcp.so) //<------
#1 dlsym() (from batch_queue.so)
#2 dlsym() (from your mpc_framework.so)
#3 pthread_create() (from mpc_framework.so)
#4 pthread_create() (from dmtcp.so)  //<------
#5 pthread_create() (from a.out)
\end{lstlisting}

This leads to a stack overflow by creating a loop.  This code seems to be
present solely for IntelMPI resolving \texttt{PMI\_Init} with
\texttt{dlsym(RTLD\_NEXT, "PMI\_Init")}.  The call is currently not compiled
conditionally.  We reported it to the
developers\footnote{\url{https://github.com/dmtcp/dmtcp/issues/604}}, but our
current workaround is simply to comment it out.  This is done in the version of
DMTCP which is bundled in MPC.

\subsubsection{High-Speed Network Support}

One of the most difficult parts of the checkpoint is the high-speed network.
Indeed, as it relies on dedicated hardware, it represents the possibility of
shared state located outside processes' memory.  As a consequence, saving
process state is not sufficient to restore connections over HPC networks such
as Infiniband or Portals.  For example, memory pinning register segments in the
device (to allow address translation and to retrieve authentication tokens) is
not checkpointable.  In order to circumvent this issue, DMTCP provided a plugin
completely wrapping the libverbs (low-level Infiniband programming interface)
in order to track and preserve a shadow state of all the operations taking
place on the card\cite{IBCR}.  This approach enabled transparent checkpointing
of Infiniband networks, but not without some drawbacks.

\begin{figure}[ht!]
	\centering
	\includegraphics[width=\linewidth]{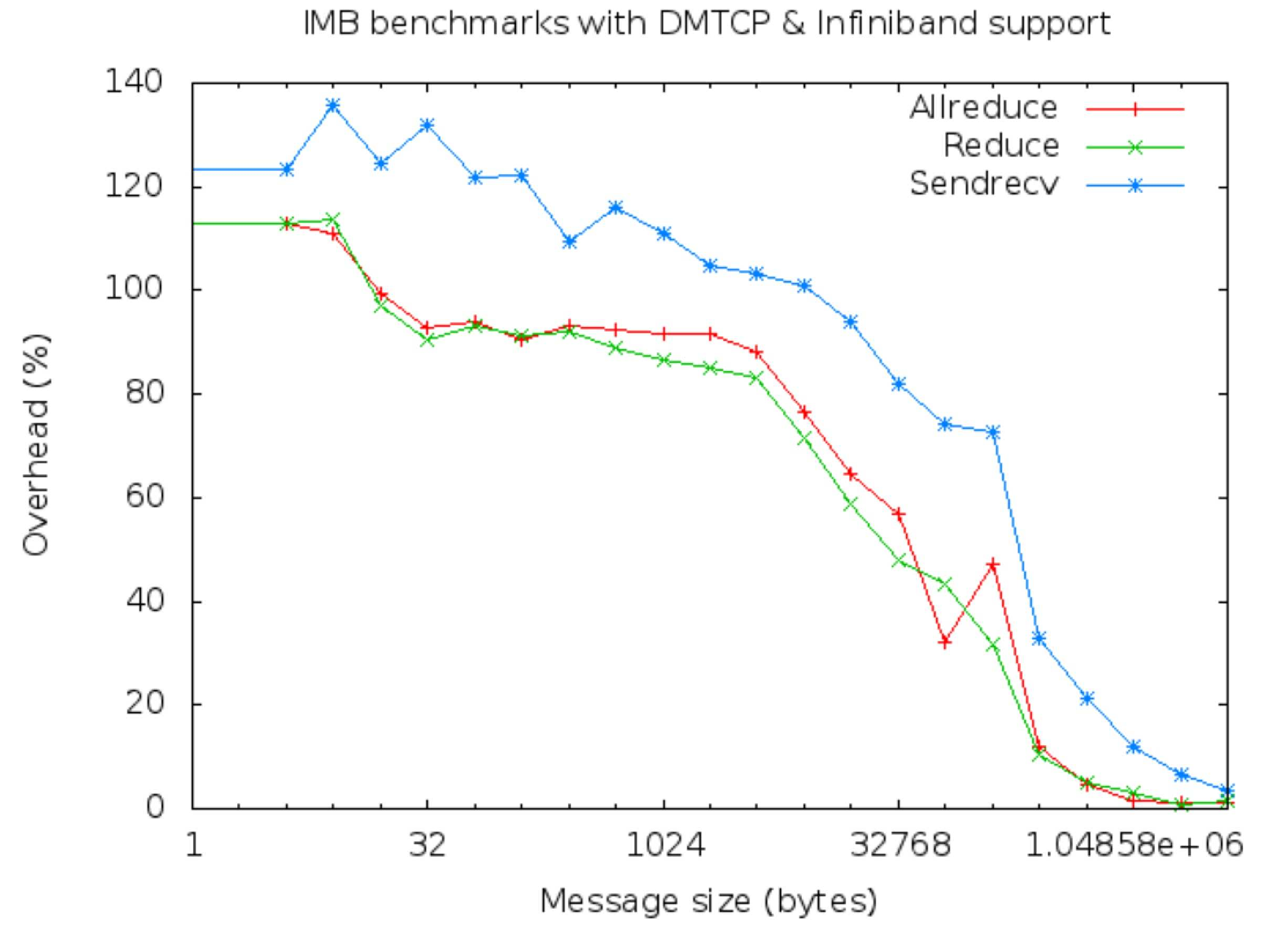}
	\caption{Overhead for Infiniband wrapping in DMTCP.}
	\label{fig:IB}
\end{figure}

As presented in Figure \ref{fig:IB}, wrapping Infiniband has a direct impact on
common MPI implementations.  Indeed, as libverbs calls are by definition on the
critical path of any IB communication, this extra wrapping leads to a
performance overhead.  We observed up to 140\% overhead for small messages,
where the extra latency is most visible.  The main drawback of the approach is
that it imposes this performance loss outside of checkpointing sections,
leading to a permanent slowdown.  It is this problem that encouraged us to look
for other alternatives mitigating the cost.

MPC's network has been built as a modular set of driver instances (called
rails) stacked on top of each other (Section~\ref{modular}).  Moreover,
on-demand connections are managed with in-band messages, which can be routed
through a dedicated signaling network (Section~\ref{bootstrap}).  Then, without
any action, routes existing prior to the checkpoint will be included in the
checkpoint, as present in internal data structures.  However, some of these
routes will be invalid at restart, because part of information they relied on
are now undefined.  While TCP network is fully handled by DMTCP with a minimum
cost, this is not the case for Infiniband.  It is not possible to purge the
multi-rail undefined endpoints efficiently after each restart because we cannot
ensure the state of the network layer when it has been stopped at checkpoint
time, potentially leading us to deadlocks.

Thus, we consider removing these routes before checkpointing the application.
Rails not checkpointable had to be fully closed each time a checkpoint is
performed.  This means that MPC frees all the resources linked to a given
driver and proceeds to remove the routes from the multi-rail lists (see Figure
\ref{fig:mr}).  Some drivers are exempted from this closing as they are
compatible with DMTCP (e.g., TCP and SHM).  In this case, static routes from
the original topology are preserved.  For these last two drivers, DMTCP will be
able to restore a state matching one of existing routes known to the process.
Dealing with drivers requiring to be closed, there will be no route associated
with these rails in the restarted process image, a new rail will be allocated
from scratch.

\subsubsection{Checkpointing Interface} \label{iface}

From an end user's point of view, this paper defines a new MPI collective
function call, whose role is to realize a transparent checkpoint.  Furthermore,
we define a set of constants linked to the state of the parallel program:

\begin{figure}[ht!]
	\centering
	\begin{lstlisting}
	int MPIX_Checkpoint(MPIX_CR_state_t* state);
	\end{lstlisting}
	\caption{Proposed transparent checkpoint interface.}
	\label{fig:iface}
\end{figure}

\begin{table}[ht!]
	\centering
	\begin{small}
		\begin{tabular}{|c|l|}
			\hline
			CR Constant & Definition \\
			\hline
			\hline
			\texttt{MPIX\_CR\_STATE\_ERROR} & An error has occurred \\
			\hline
			\texttt{MPIX\_CR\_STATE\_CHECKPOINT} & The program has checkpointed \\
			\hline
			\texttt{MPIX\_CR\_STATE\_RESTART} & The program has restarted \\
			\hline
			\texttt{MPIX\_CR\_STATE\_IGNORE} & Command ignored (not supported) \\
			\hline
		\end{tabular}
	\end{small}
	\caption{\texttt{MPIX\_Checkpoint} constants definitions.}
	\label{tab:meanings}
\end{table}

As presented in Figure~\ref{fig:iface}, the \texttt{MPIX\_Checkpoint} call is a
collective with respect to \texttt{MPI\_COMM\_WORLD}.  It will return to a
state defined in Table \ref{tab:meanings}. Entering this function means the
application is requiring the MPI implementation to create a new checkpoint,
there should be no unmatched MPI messages to prevent message losses.  One point
to note is that this call can return in different scenarios.  First, a program
returning from a regular checkpoint proceeds to call \texttt{MPIX\_Checkpoint}.
In this scenario, the function call will return each time \texttt{CHECKPOINT}
when the step completes.  When the application program restarts, the work-flow
will immediately come from \texttt{MPIX\_Checkpoint} and the return value will
be \texttt{RESTART}, allowing the application to be notified of the current
state (post-checkpoint or restart). If it is not possible to checkpoint (e.g.,
due to lack of support), a runtime can return \texttt{IGNORE} to inform the
application that nothing was saved.

The collective nature of the call also ensures that it is correctly invoked in
the case of a hybrid program.  For instance, if this function is called in an
OpenMP parallel region, it will require the application to implement a critical
region so as not to violate the collective nature of the call.  By clearly
stating how the checkpoint function is to be called globally, it abstracts the
integration of such a call, while simplifying the implementation requirements.

\subsection{Evaluating our DMTCP Integration}
\label{issues}

It is important to observe that a direct consequence our transparent
checkpointing approach is that it closes dynamic routes at each checkpoint.
Indeed, in order to create a valid process image, we alter the state of the
application even if it does not go through a restart. This is currently a
limitation of our model as later communications will immediately recreate
routes previously closed for the sole purpose of a checkpoint. Initially, we
envisioned to simply remove uncheckpointable endpoints from the multi-rail list
(see Figure \ref{fig:mr}) without freeing any memory. This, however, led to
various issues, first obviously a memory leak with the added complexity that it
was not possible to free this dangling memory at the restart. Second, leaving
an open device, for example, the IB HCA, means that there is an open file
descriptor upon checkpoint that DMTCP will try to drain, eventually leading to
a deadlock. Consequently, dynamic route closing and its associated performance
impact appeared to be a good tradeoff in the case of Infiniband networks. Other
network types, in particular, connection-less networks such as Portals 4 or
Omnipath, may circumvent this limitation. We are currently studying this
possibility.

\begin{figure}[ht!]
	\centering
	\includegraphics[width=\linewidth]{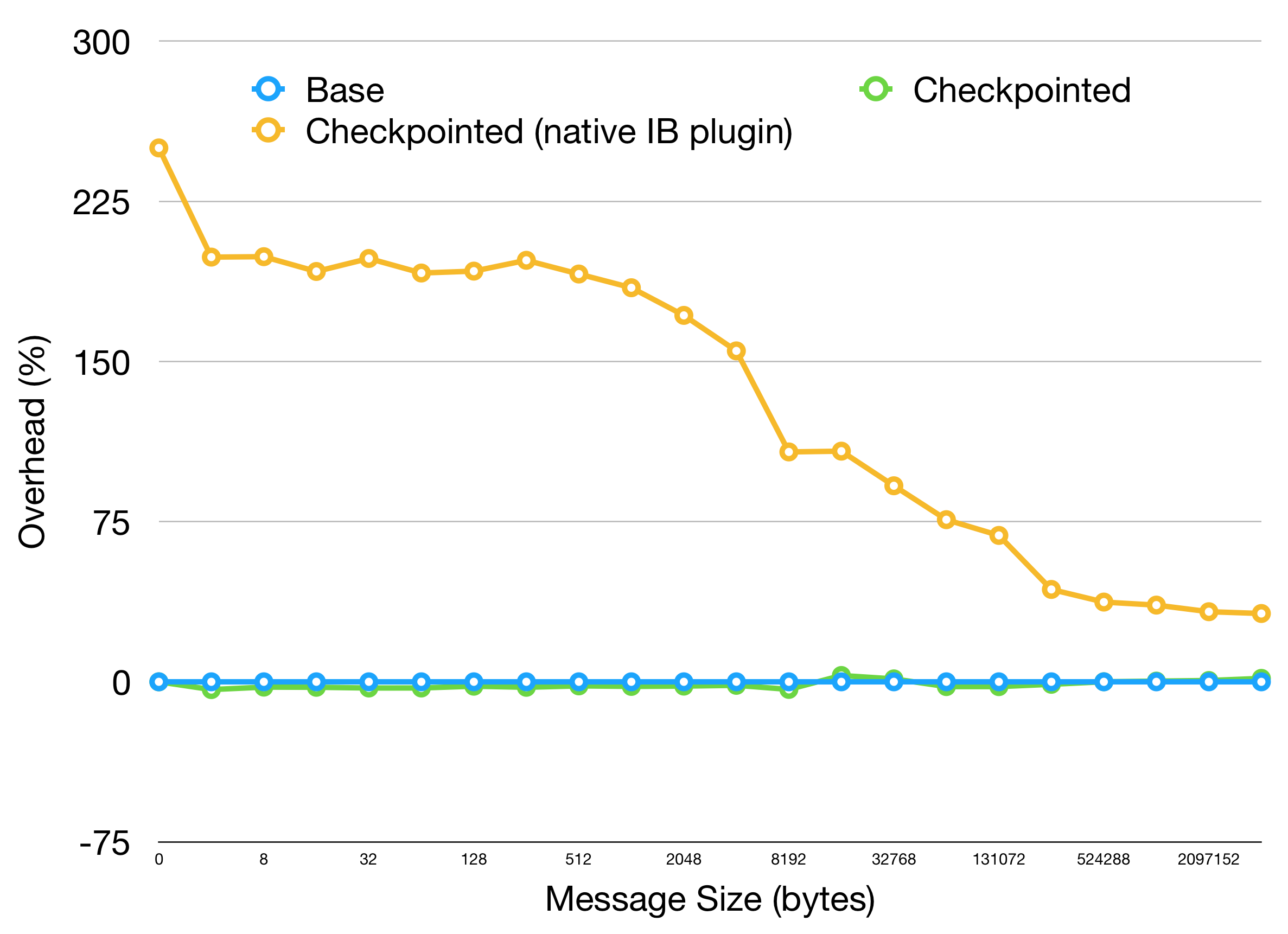}
	\caption{IMB Allreduce performance overhead between DMTCP Infiniband support and MPC's support.}
	\label{fig:imb}
\end{figure}

Figure \ref{fig:imb} compares the performance of our high-speed network
checkpointing methodology to the Infiniband wrapping one. A direct execution
shows no measurable overhead on communications when starting from a checkpoint
on the IMB benchmarks and the restarted program has similar performance to the
initial process image. The reason is because the only penalty taken by the
restarted program is route creation which is a punctual process mitigated by
the repetitive communication pattern in communications. The checkpoint by
itself, however, has a performance cost as it closes connections, nonetheless,
we believe that this is an acceptable point as the user is free to set its
frequency. In summary, our method creates a transient overhead to prevent a
permanent one.

The duration of a given checkpoint is highly dependent on both the scale of the
MPI job and the amount of memory it uses.  Moreover, the wall-time overhead it
incurs is correlated with the number of checkpoints performed during a given
execution.  Consequently, as for checkpointing, we are willing to leave the
application untouched, the only parameter available to limit the overhead is
checkpoint frequency. If we consider a computation lasting $T_s$ seconds and a
checkpointing time of $T_c$ every $\tau$ second, we have the following total
duration $D$ of the checkpointed program: $D = T_s + \frac{T_s}{\tau}T_c$.
Denoting $f = \frac{1}{\tau}$ as the checkpointing frequency, we immediately
have $D = T_{s}(1 + fT_c)$.  Now reasoning in terms of overhead, we have
$O_{vh} = \frac{D}{T_s} = 1 + fT_c$, this shows that the overhead is
necessarily positive and easily computable from both checkpoint frequencies and
duration.  More importantly, it can easily be budgeted.  For example,
considering a one-minute checkpoint time and a maximum overhead of 1\% we have
$f = \frac{1\%}{T_c}$ and therefore a checkpointing period $\tau = 6000$
seconds or 1 hour and 40 minutes.  This small formula shows that it is
relatively easy to amortize the checkpointing time through the frequency
parameter in a reasonable time.  When measuring the Intel Messaging Benchmark
(IMB), we encountered checkpoints around three seconds for 32 MPI processes --
$T_c = 60$ is then already a pessimistic value.

\subsubsection{DMTCP Performance Evaluation on Lulesh}

In order to assess the performance of our checkpointing methodology, we ran it
at scale on the Lulesh\cite{LULESH} benchmark. In particular, we focused
ourselves on two aspects. First, the checkpoint time that can be directly
connected to a global overhead given a checkpointing frequency -- as outlined
earlier. Second, we want to measure the cost associated with closing
connections in terms of execution time outside of checkpoints. Measurements
were carried over on a small test system at CEA featuring Sandy-Bridge
processors and 16 cores per node using a problem size of 30. Interconnect
consists in mlx4 Infiniband Host-Channel Adapters. In order to characterize the
cost of our methodology, we proceeded to measure a single checkpoint in the
middle of the parallel execution at various scales. Lulesh was launched with a
single process per node in MPI+OpenMP configuration. Runs were done with a
fixed size of 30 (\texttt{-s} flag), given Lulesh's design this size is given
as the size per process and therefore our measurements are all done in a weak
scaling fashion -- problem size is $30^3$ per MPI process.

\begin{figure}[ht!]
	\centering
	\subfigure[Walltime breakdown in seconds]{
		\includegraphics[width=\linewidth]{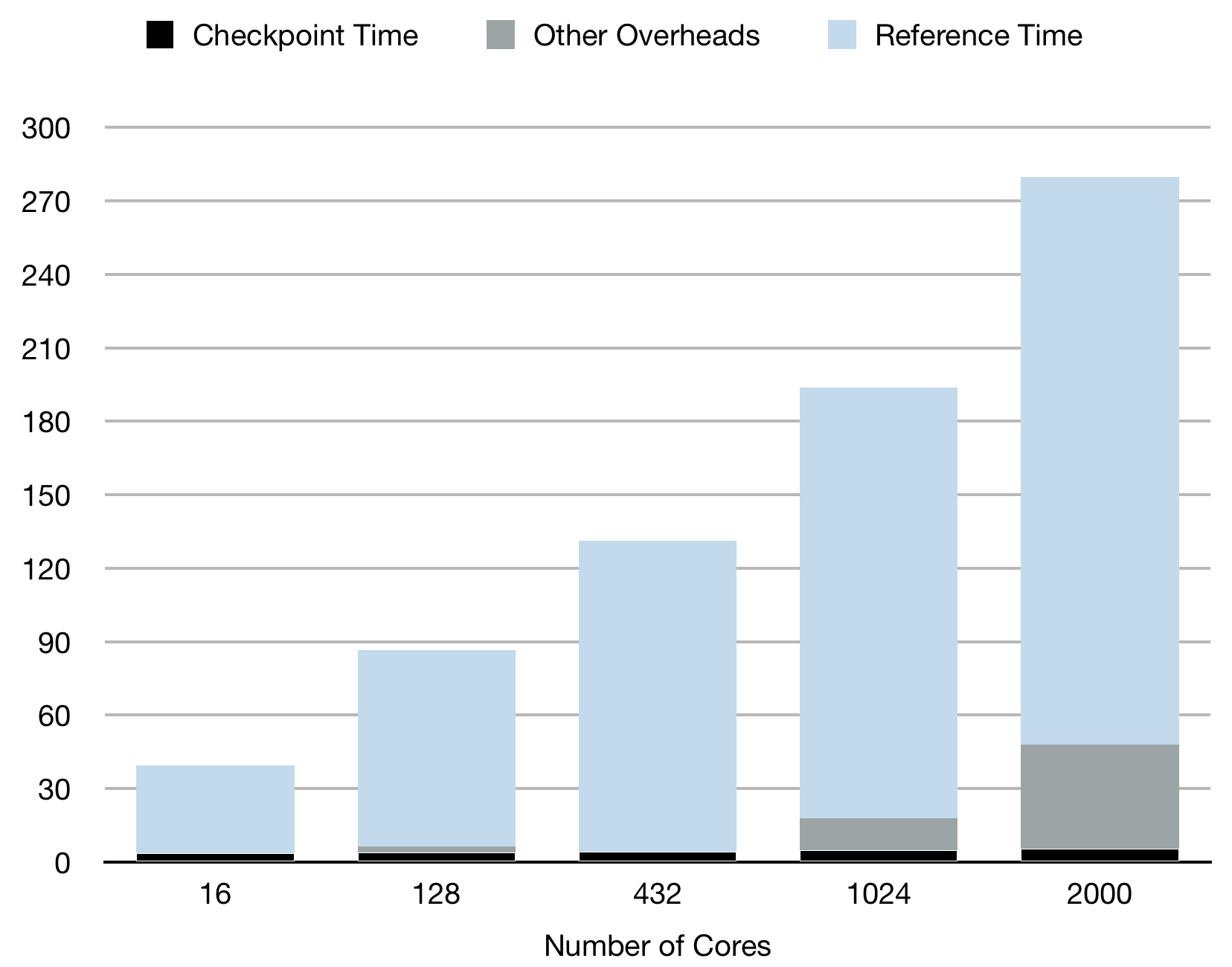}
		\label{lultime}
	}
	\subfigure[Walltime breakdown in percentage]{
		\includegraphics[width=\linewidth]{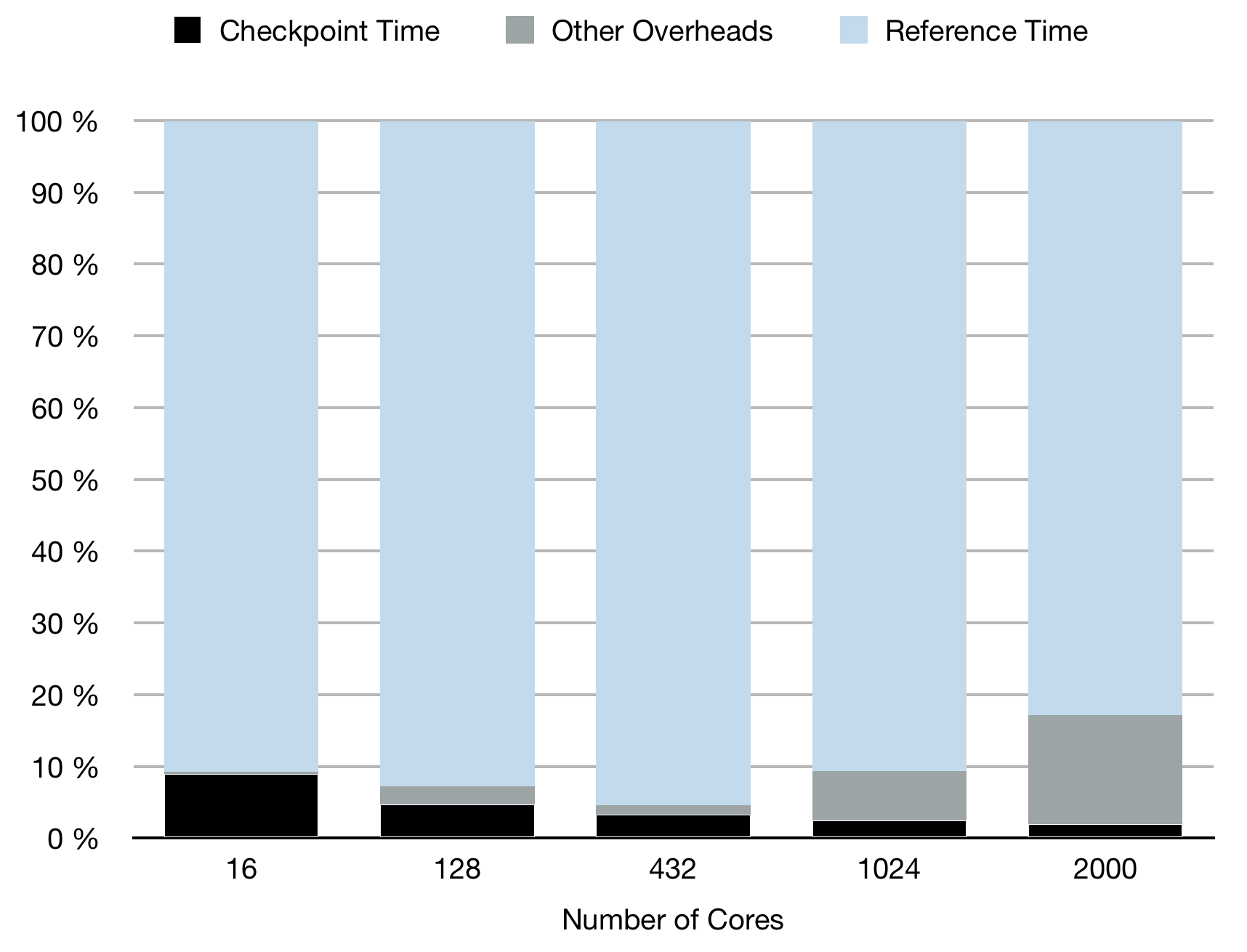}
		\label{lulovh}
	}
	\caption{Checkpoint and reference times for Lulesh (size 30) in function of the number of cores.}
	\label{lulrez}
\end{figure}

In Figure \ref{lultime}, we see the breakdown of the walltimes in terms of
reference time, checkpoint and other overheads. The checkpoint overhead is the
time spent generating the data in the collective call, other overhead accounts
for other differences with respect to reference time, including on-demand
connections. In Figure \ref{lulovh}, we present the same results as a
percentage, to highlight the relative cost of each time. What can be seen that
that the checkpointing time by itself remains relatively steady as the number
of cores increases. However, we observe a rise in indirect costs from 0.3 \% up
to 18 \% when considering large scale. This can be explained by the high number
of on-demand connections, as the number of nodes gets larger. This overhead,
mitigating the I/O saturation effect linked with an increasing number of MPI
processed saving their state in parallel is also correlated to the transitive
cost of disconnecting and then reconnecting MPI processes. This cost is then
clearly not negligible. However, and as we are now going to discuss, the nature
of the checkpoint and in particular its punctual nature can be used to slightly
reduce performance impact.

\subsubsection{Checkpointing Lulesh with Constant Overhead}

\begin{figure}[ht!]
	\centering
	\includegraphics[width=\linewidth]{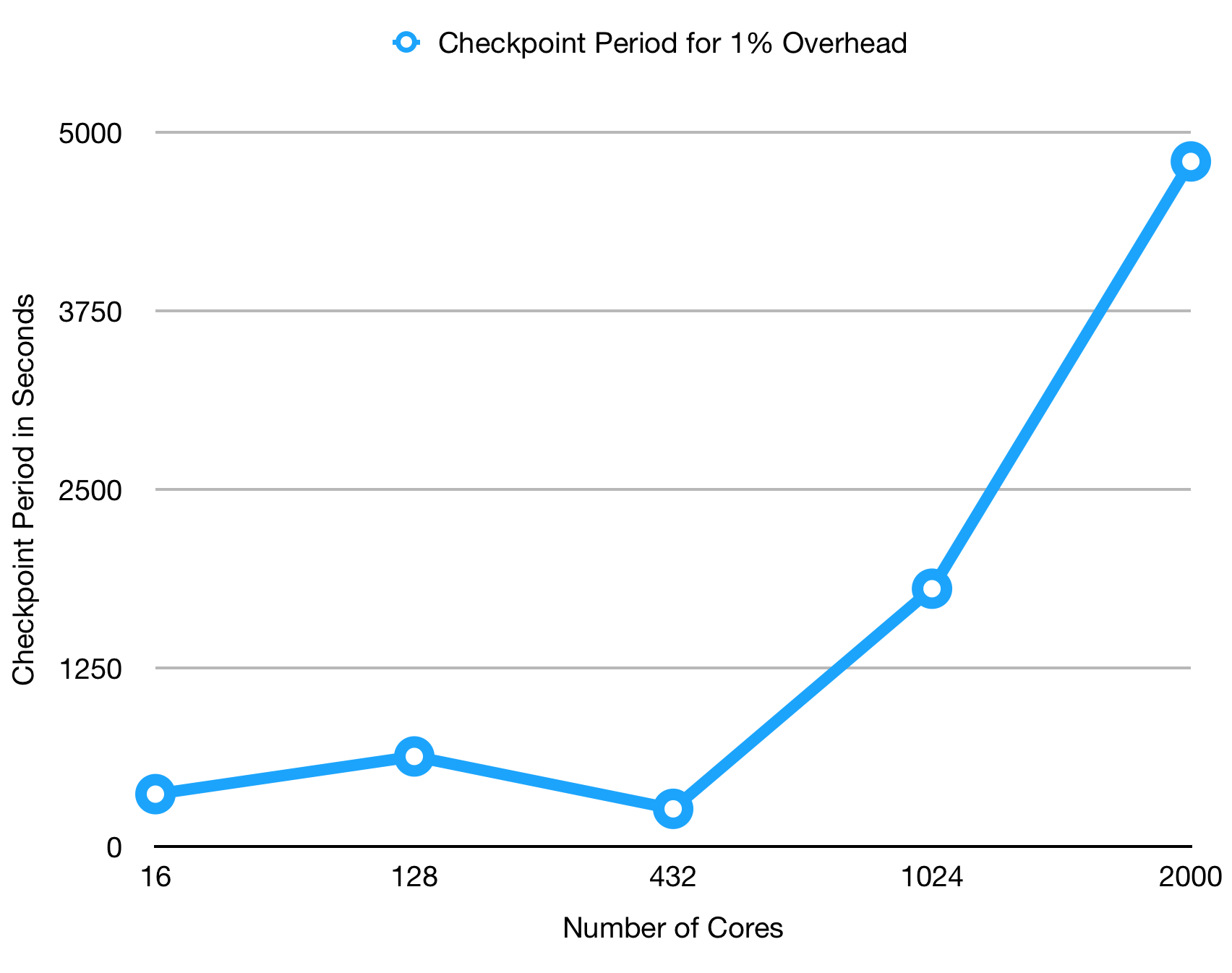}
	\caption{Checkpoint period for 1\% overhead computed from results presented in Figure \ref{lulrez}.}
	\label{ckovh}
\end{figure}

In order to expose these results in a more practical manner, we used the
formula presented in the previous section to compute the checkpointing period
such as the overhead is 1\% in the light of previous measurements. To do so, we
added direct and indirect overheads, considered  as the total checkpoint cost.
This yielded the values presented in Figure \ref{ckovh}. One can see that
despite potentially expensive, the checkpoint cost on the walltime can be
mitigated for long-running programs. In our case, for Lulesh, we see
checkpointing periods ranging from five minutes at one process to one hour and
twenty minutes at larger sizes. The overall checkpoint cost rapidly increases
requiring checkpoints to be further spaced to mitigate their apparent cost.
This can find its explanation in several factors that we described as
\emph{other overheads} in Figure \ref{lulrez}. Indeed, the number of
connections to be restored is dependent from communication topology which is
polynomial in the case of Lulesh's 3D mesh. In addition, the saturation effect
on I/O caches and more generally the file-system is not to be neglected as the
number of processes increases. These two factors are possible explanations to
the important overhead as the number of cores increases. However, we think that
our methodology is still usable at scale as checkpointing periods of a few
hours are not unrealistic. Eventually, it is important to note that the cost is
directly linked not only to scale but also the data-set  size manipulated by
the application -- transparent approached dumping full process images. Dealing
with this later constraint, application-level checkpointing has a lot of
advantages as it can benefit from application developer's input at the
impediment of the associated programming cost.

\subsubsection{Generalizing to other MPI Implementations}

Results presented in this paper were obtained with the MPC runtime which
provides support for checkpointing. In particular, we presented a dedicated
collective call \texttt{MPIX\_Checkpoint} and relied on high-speed network
disconnection prior to checkpointing. In addition, MPC include the signaling
network which can be restored by DMTCP (being in TCP) and latter used to relay
on demand connection demands to reconnect routes. This second aspect is not
compulsory to enable checkpointing and therefore allows our methodology to be
adaptable to other runtimes. Compulsory requirements are (1) the ability to
close high-speed connections and to restore them later on and  (2) the capacity
of restarting either from the PMI or using a support network (one may consider
launcher processes). This methodology can then be adapted to other runtimes and
is not dependent on MPC, it simply requires state management capacities in MPI
for connections and startup -- DMTCP handling most of the checkpoint.

\section{Application-level Checkpointing}

In this second part we focus ourselves on how checkpointing could be achieved
at application-level. In previous Section we have seen how it was possible to
save the distributed state of an MPI application without requiring substantial
modifications, just a single line of code at time-step level. However, and as
we also commented this method despite practical is far from optimal as it
supposes the saving of more than actually needed by the application. This leads
to inefficient checkpoints in terms of memory and overhead linked to the
associated I/Os. This second approach requiring application developer input is
then more efficient in terms of storage space, although it requires the program
state to be fully serializable. This supposes that the state of each library
can be correctly intercepted which is sometimes not practical, considering for
example third-party software. In this Section, following  what we have already
done for DMTCP, we will first introduce the FTI library which aims at exposing
convenient mechanism for application-level checkpointing. Then, we present our
integration inside MPC. We show some performance results in a heat-dissipation
benchmark which was easily ported to FTI. We also ported Lulesh on top of FTI
to enable further comparisons with the transparent method. This comparison,
however, has to be mitigated as measurements were made on a different machine
due to organizational constraints.

\subsection{FTI Overview}
\label{sec:2-fti}

FTI is a multilevel checkpointing library with a wide set of features. The
purpose of this library is to provide an interface to address the various
storage levels in high-performance computing environments for checkpointing
purposes.

High-performance computing is an always evolving field in which new hardware
devices are continuously being developed and integrated; not only for computing
but also for storage. As a consequence there exists a discrepancy between types
of storage in terms of performance, availability and reliability. For instance,
mechanical hard-disk drives usually offer higher capacity but lower performance
than solid state drives. Such trade-offs are at the core of the concept of
multilevel checkpointing, with the goal of finding the sweet spot in the
reliability versus performance trade-offs. To further illustrate this, FTI
offers the following four checkpointing levels:

\begin{enumerate}[Level 1.]
	\item Checkpoint in local storage.
	\item Local checkpoint and copy on a partner node.
	\item Local checkpoint with erasure coding.
	\item Checkpoint in the PFS.
\end{enumerate}

Level 1 checkpoint is the least reliable level but also the fastest, while
Level 4 is the most reliable but also the slowest of all levels. Given that
most failures in supercomputers do not affect all nodes simultaneously, there exist
possibilities to combine specificity from each level to yield improved performance,
this is the goal of the FTI interface that we describe in the next Section.

\label{subsec:2.2-fti_usage}

Writing checkpoints in local storage is sufficient to put up with soft errors
but cannot withstand node failures, data stored in the local storage being
inaccessible until the node is repaired. Therefore, local checkpointing has to
be combined with some sort of data redundancy in order to tolerate one or
multiple node crashes. FTI implements several approaches for this purpose, such
as data replication on a partner code, data redundancy through Reed-Solomon
encoding or data persistence into the parallel file system.  Application-level
checkpointing thanks to its improved data selectivity is an efficient method in
terms of minimizing data to be saved. In addition, addressing different types
of storage (e.g. node local) further enhance performance. Nonetheless,
checkpointing remains an expensive operation. Therefore, it can be optimized by
dividing it into two stages: (1) write the checkpoint in local storage and (2)
apply data redundancy in the background while the application continues its
execution.

Data redundancy techniques systematically require some sort of processing,
either by transferring data through the network or by performing extra
computations. This additional work can be done locally, impeding extra overhead
to the application or using dedicated hardware (like a RAID array).  It is that
capacity we want to provide more efficiently thanks to the runtime support.

\begin{figure}[htb]
	\centering
	\includegraphics[width=\columnwidth]{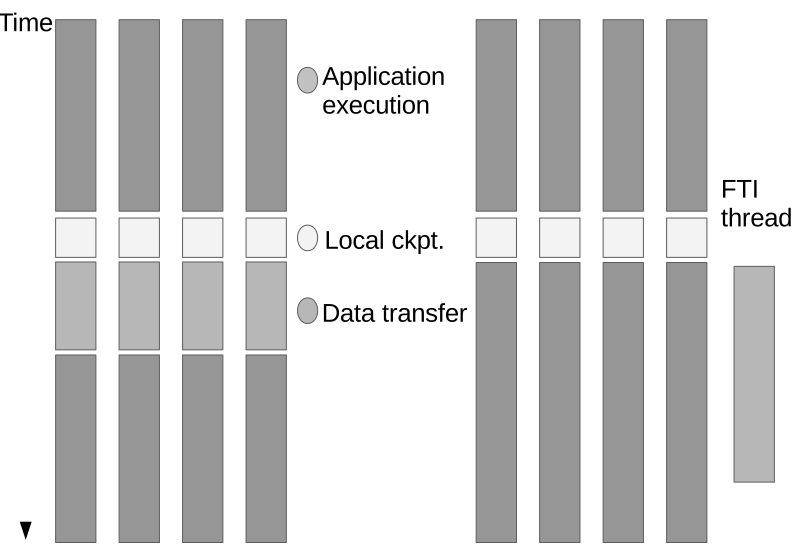}
	\caption{FTI synchronous vs asynchronous transfer}
	\label{fig:fti-async}
\end{figure}

\label{subsec:2.3-fti_dedicated}

As far as this post processing is concerned, FTI offers the option to ``steal''
one process per node from the application in order to perform these tasks.  In
this case, the dedicated processes are isolated from the application processes
by splitting the global communicator and providing a new one to the
application.  The FTI helper processes then run in their own communicator to
perform their resiliency-related tasks.

Using this technique, application processes can pursue their execution as soon
as the local checkpoint has been made. Data replication is offloaded to these
helper processes in parallel to the regular execution as shown in
Figure~\ref{fig:fti-async}. This has been proven to be quite efficient, because
it virtually transforms all checkpoints into local checkpoints. Unfortunately,
this supposes the use of dedicated resources for this purpose. Moreover, most
large scale supercomputers do not allow applications to run more processes than
there are on a given node due to batch-manager constraints. For this reason we
explored the possibility of relying on oversubscription in MPC's user-level
scheduler to perform such tasks in the background without dedicating specific
resources.

\subsection{MPC's Unified user-level thread scheduler}
\label{subsec:3.2_unified_sched}

Before explaining how we integrated FTI in MPC, this Section recalls some
aspects of MPC's scheduler. First, it handles user-level threads, bypassing the
OS scheduler, often ill-suited for HPC parallel applications.  An MxN
user-level scheduler as in MPC bypasses the OS scheduler, one OS thread is
created per computing unit on a node, and pinned to a given computing unit. On
top of this thread, the user-level scheduler handles the selection of user
threads to be executed on this OS thread. This way, scheduling decisions,
previously delegated to the OS scheduler, are now handled by the user-level
threads in-place, as there are as many OS threads than cores. Scheduling
policies are then completely handled by the user-level thread scheduler.

By collocating multiple programming models in its user-level scheduler, MPC is
able to coordinate the execution of threads from various origins. With its
global view of the whole node, all available computing units and threads to be
executed, the scheduler can then make the best decision possible according to
the scheduling policies. This feature could be illustrated with
oversubscribing. Indeed, since everything is a thread the scheduler knows when
a thread is idle and may replace it with another active one. In the case of
oversubscription, it means that as soon as a thread is idle, an extra thread
can be scheduled to use these idle resources. This thread may originate from
any of the active models. This helps maximizing the usage of available
resources, and reduces performance loss, due to waiting and idle threads. It is
this last propriety that motivated the integration of FTI inside MPC, exploring
the possiblity of collocating the \emph{helper} process in an oversubscribed
thread instead of a POSIX process.

\subsection{Supporting FTI in MPC}
\label{sec:4-contrib}

Now that we presented our integration of DMTCP inside MPC, this Section focuses
on the integration of application-level checkpointing in the context of the FTI
library. As explained in previous sections, MPC provides its own implementation
of the MPI standard. Since FTI is relying on MPI to implement its
checkpoint/restart method, we simply used MPC as the MPI implementation for
FTI.  The rest of this section, we first detail the port of FTI atop MPC and
then describes how checkpoint data post-processing took advantage of MPC's
user-level scheduler and oversubscription.

\subsubsection{Port of FTI atop MPC}
\label{subsec:4.1-fti_mpc}

This step was simple to achieve. As MPC is an MPI implementation, the only
necessary action was to compile FTI with MPC compiler wrappers
(\texttt{mpc\_cc}) relying on ``automatic privatization''.

\label{subsec:4.32-mpc_oversubscribe}

After checking that FTI and MPC were correctly collaborating both in
process-based and thread-based configurations, we sought to benefit from MPC's
specificities. In particular, we moved the dedicated MPI process inside a
user-level thread to benefit of oversubscription.

In Section~\ref{subsec:2.3-fti_dedicated}, we explained how having a dedicated
MPI process for post-checkpoints per node improves performance.  However, this
is mainly true when the MPI process runs on its own resources.  If no core is
available, this additional MPI process will be oversubscribed.  It means that
this additional MPI process shares resources with the original application. As
two processes, or even threads, cannot run at the same time  on the same core,
their respective code will be executed, turn by turn, after context switches.

In order to mitigate oversubscription overhead, we targeted MPC's thread-based
MPI capabilities. The interest is twofold with (1) lighter context switches and
(2) the ability to use MPI waiting time (in the application) to progress
checkpointing. We saw in Section~\ref{subsec:3.2_unified_sched} that the MPC
scheduler uses its own user-level threads. Hence, in a full thread-based mode,
each MPI process on a node is a user-level thread managed by a unified
scheduler.  Switching from one MPI process to another is ``just'' a user-level
thread context switch, which is lighter than between two UNIX processes.  This
then makes the approach involving an oversubscribed MPI process more attractive
than in a regular process-based MPI setup.

\subsection{Evaluating the FTI Integration}

Now that we presented the results with  user-level transparent checkpointing,
this Section now studies the impact of our MPC integration on FTI for
application-level checkpointing.  In particular, we compare performance between
additional MPI processes and our oversubscribed model, taking advantage of
user-level threads.

In a first approach, we ported Lulesh on FTI.  To do so, the first step was to
change all \texttt{MPI\_COMM\_WORLD} references to \texttt{FTI\_COMM\_WORLD}.
If such replacement can be tedious on large production code, the LULESH code
infrastructure simplified this change as all MPI calls are located in only two
files.  The next phase to port an application to FTI relies on the data
election to be saved for checkpoint. Here again, the code structure was
conveniently outlining important data in a single C++ class. However, one
highlighted issue by porting LULESH to FTI was this C++ class. Indeed, FTI can
handle C structures natively with \emph{FTI\_InitType()}. However, this
checkpointed dataset includes a tree-like structure relying on pointers. As
such pointer cannot be saved due to memory remap upon restart, we had to
serialize the structure.  As a consequence, this C++ class object is serialized
into a buffer, handled to FTI. Thus, before checkpointing, this serialization
had to be performed.  Symmetrically, this buffer is de-serialized into the C++
object upon recovery. The BOOST library was used handle the serialization.  The
port of LULESH to FTI, and the serialization process were validated by
arbitrarily killing the job at different times before recovering the job and
resuming execution.  All program outputs remained valid whatever the number of
MPI processes (up to 1728).

\begin{figure}
	\centering
	\includegraphics[width=0.45\textwidth]{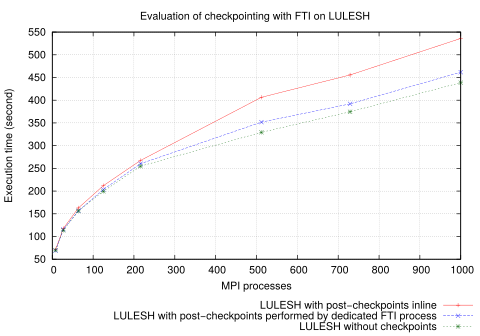}
	\caption{Performances without and with FTI checkpointing methods, no oversubscribe}
	\label{fig:fti_impact}
\end{figure}

\begin{figure}[ht!]
	\centering
	\includegraphics[width=\linewidth]{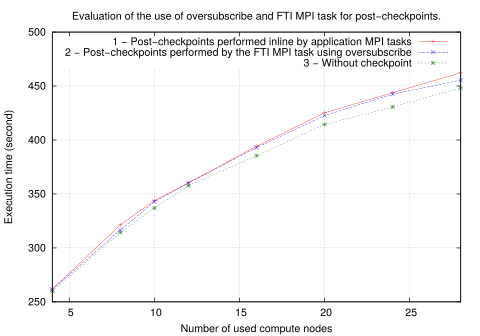}
	\caption{Oversubscribing with an MPC MPI thread}
	\label{fig:mpc_oversubscribe}
\end{figure}

As presented in Figure \ref{fig:fti_impact} which does not rely on
oversubscription, using a dedicated checkpointing process is advantaging when
compared to the inline approach which does not provide any overlap. This shows
that there is an interest in integrating such support though an user-level
scheduler. Since Lulesh works with numbers of MPI processes which are power of
3, it was not possible to produce a sufficient number of  configuration where
all cores are loaded with computation, and thus to realize oversubscribing when
using a dedicated FTI process for the checkpoints. We tested our oversubscribed
application-level checkpoint restart approach on a heat distribution benchmark
(heatdis). Heatdis is a 2-dimensional stencil code that distributes a 2D grid
among MPI processes. Processes only communicate with neighbor processes for
exchanging ghost cells. As this benchmark does not impose restriction on the
number of MPI processes (unlike Lulesh), we were able to validate multiple
configurations. Performance measurements were realized on the MareNostrum 3
supercomputer at the Barcelona Supercomputing Center (BSC).  MareNostrum 3 is a
1.1 petaflop peak performance supercomputer with Intel SandyBridge processors.
The machine features 3056 nodes connected through an Infiniband FDR network.
As presented in Figure \ref{fig:mpc_oversubscribe}, we ran this benchmark in
different configurations:

\begin{itemize}
	\item without FTI to provide a base time;
	\item with FTI and inline post-processing;
	\item with FTI and a dedicated oversubscribed MPI process (running as thread).
\end{itemize}

It can be seen that when relying on a thread to perform post-checkpointing
operations the overhead is slightly lower than if it was done inline, directly
impacting the code. This shows that such a model can lead to some benefits when
being used in threads. However, gains are still relatively limited. We think
that the non-preemptive nature of the scheduler and the fact that our
integration of file I/Os in MPC is not fully taking advantage of the scheduler,
preventing MPC from inserting yielding points inside I/O operations (when being
blocked in a \texttt{write}, for example). Indeed, integrating a model inside
user-level threads requires generally a complete wrapping of every call to
avoid cases potentially blocking the OS thread carrying the execution. For this
reason, mutexes, semaphores, Pthread operations, and so on, are captured by MPC
to be managed accordingly in the unified scheduler. We think that converting
I/Os to non-blocking operations and accounting for it in the scheduler as yield
points should bring improved performance compared to what is presented in
Figure \ref{fig:mpc_oversubscribe}.

\begin{figure}[ht!]
	\centering
	\includegraphics[width=\linewidth]{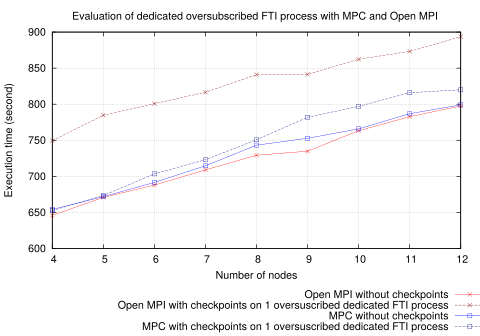}
	\caption{Comparing oversubscribe with MPC and OpenMPI}
	\label{fig:mpc_ompi}
\end{figure}

However, for users willing to fully use all cores by oversubscribing the FTI
processing instead of inlining it, performance gains can be obtained with
user-level threads.  Indeed, as depicted in Figure \ref{fig:mpc_oversubscribe},
the same benchmark has been run with MPC (user-level threads) and OpenMPI (UNIX
processes) in order to compare oversubscription costs.  Since the previous node
configuration was not available during these tests, the comparison has been
realized on a Sandy Bridge partition, with 16 cores per node.The Heatdis
benchmark has been initially run on both MPC and OpenMPI, with disabled
checkpointing, yielding to similar results. However, as far as oversubscription
is concerned, it can be seen that user-level threads yield lower overhead than
processes --  OpenMPI showing an additional overhead between 10\% and 15\%,
depending on the number of nodes.  One of the main reasons for this is that MPI
is designed to efficiently poll the resources of its core to progress
communications with minimum latency whereas in MPC, this additional process
directly benefits from the existing shared-memory communication layer. This
allows the helper MPI processes to yield to a computation process in a fairer
manner, limiting the overhead.

To summarize, the use of FTI inside user-level threads shown that
oversubscription was more efficient than with OS processes. Reasons for this
are the unified communication layer mitigating potentially aggressive polling,
leading to improved fairness between threads and more efficient context
switches. However, we observed reduced gains when it comes to I/O integration
in the scheduler as we did not integrate yield points inside POSIX I/Os in
MPC's non-preemptive scheduler, mitigating the potential overlap of an I/O
intensive thread such as the one exposed by FTI. We are considering to address
this issue as future work.

\section{Conclusion}

The paper presents our implementation of transparent checkpointing in the MPC
MPI runtime. Based on our knowledge, it is the first illustration of
transparent checkpoint restart -- agnostic from the application -- with network
support in a thread-based MPI. Checkpointing has already been illustrated in
runtimes involving user-level threads in the past, like Charm++\cite{CHARM} and
its combination with AMPI\cite{CHARMAMPI}.  Our approach is more general as it
does not rely on serialization assumptions in terms of application's
programming model, aspect directly inherited from DMTCP's versatility. However,
as we put no constraints on the application, some scenarios possible with
Charm++ are out of reach, they include in-memory
checkpointing\cite{CHARMMEMORY} and restarting the program on a different
number of processes\cite{CHARMMIG}.  In our case, we solely presented a
synchronous checkpointing interface which is only a subset of what is possible
in terms of fault tolerance.  Indeed, new interfaces such as ULFM in MPI should
allow applications to react to failures at runtime -- limiting the need for
restarts from scratch, as provided by DMTCP.  Moreover, our approach does not
support partial checkpoint restart, it is nonetheless a point that we would
like to explore in the future.

In addition, we focused ourselves on application-level checkpointing with the
help of the FTI library which targets multi-level checkpointing by providing
application developers with a dedicated checkpointing API.  The approach has
the advantage of benefiting from developers' knowledge to limit the checkpoint
size when compared to transparent approaches. However, it has the drawback of
requiring modifications in the target application.  Consequently, despite
yielding the same checkpoint/restart result, the application-level model has
different implications and can then be seen as complementary to transparent
methods. Our integration in the MPC runtime relied on user-level threads to
perform post-checkpointing processing (data replication) and demonstrated that
in terms of oversubscription, where MPI processes running in threads were more
efficient than those using regular UNIX processes. Although,  MPC's scheduler
is lacking when it comes to handling blocking I/Os in a non-preemptive manner,
we would like to address in the future.

In this paper we presented two approaches for checkpoint-restart in the context
of a thread-based MPI called MPC. In particular, we considered two different
kinds of approaches at application- and user-level and discussed how they
collaborated with our runtime either at communication level or with a unified
scheduler. This showed that runtimes can provide mechanisms to improve
checkpointing efficiency and that such mechanisms were applicable in a
relatively straightforward manner to the specificity of a thread-based MPI
runtime.

\section{Future Work}

We see several tracks of enhancements following this work. As far as the FTI
model is concerned, thread migration would allow oversubscribed MPI processes
to take advantage of idle time more efficiently as they are currently stuck in
the scheduling list of a single OS-level thread.  In addition, the integration
of I/Os in the scheduler using non-blocking file descriptors and a wrapping of
the POSIX I/O interface should improve performance when it comes to
oversubscribed I/O intensive payloads.

Regarding transparent C/R, current implementation in MPC has been designed to
provide an initial support saving our users from the development of their own
solution. However, checkpointing and more generally fault tolerance, for
example through ULFM, allows a much wider range of scenarios. Indeed, our
runtime has to fully restart in order to recover from a single node failure.
The overhead currently impacts both checkpoint and restart phases. Closing the
network could be considered a waste of time if no failure occurs between two
checkpoints. Another idea would be to save the full network structure --
updating DMTCP accordingly to disregard such network -- and paying the price of
cleaning it only at restart.

We would like to explore partial checkpointing with spare nodes leveraging our
signaling network. Another aspect that seems promising is the exploration of
connection-less networks and how they might be checkpointed more efficiently
than by actively disconnecting-reconnecting peers. In particular, we are
considering the Bull Exascale Interconnect (BXI)\cite{BXI} Portals 4
network\cite{PORTALS4} to develop such support.

\section*{Acknowledgements}

This research has been partially sponsored by the European Union’s Horizon 2020
Programme under the LEGaTO Project (\url{www.legato-project.eu}), grant
agreement 780681 and the Mont-Blanc2020 project, grant agreement n. 779877.

\appendix

\section{Usage Example}
\label{example}

In this Appendix, we give a quick overview of how to launch MPC with
transparent checkpoint-restart support. First, you need to install the last
release of MPC with \texttt{--enable-mpc-ft} option. This should install DMTCP
and enable its support in the code. When you proceed to launch the code with
\texttt{mpcrun} you may pass the \texttt{--checkpoint} option in order to
enable DMTCP's preloading and provide either checkpointing capabilities through
the coordinator or via the \texttt{MPIX\_Checkpoint} call. Eventually, to
restart a checkpointed program you may simply use the \texttt{mpcrun}  command
with the \texttt{--restart} option, taking as an optional argument the path to
the restart script generated during the checkpoint (current directory by
default). Eventually, we recommend relying on the Slurm launcher as it is the
most widely supported by DMTCP.

\bibliographystyle{elsarticle-harv}
\bibliography{sigproc} 

\end{document}